\documentclass[
reprint,
nofootinbib,
 amsmath,amssymb,
 aps,
]{revtex4-2}

\usepackage{graphicx}% Include figure files
\usepackage{dcolumn}% Align table columns on decimal point
\usepackage{bm}% bold math
\usepackage{amsthm}
\usepackage{xcolor}
\usepackage{physics}

\theoremstyle{definition}

\usepackage[colorlinks=true,allcolors=blue]{hyperref}

\newcommand{\R}{\mathbb{R}}
\DeclareMathOperator{\Span}{span}
\DeclareMathOperator{\cl}{cl}

\begin{document}

\preprint{APS/123-QED}

\title{
Gardner volumes and self-organization in a minimal model of complex ecosystems}

\author{Frederik J. Thomsen}
 \email{f.j.thomsen@tudelft.nl}%
\author{Johan L.A. Dubbeldam}%
 \email{j.l.a.dubbeldam@tudelft.nl}
\affiliation{%
 Delft Institute of Applied Mathematics, \\ 
 Delft University of Technology, Mekelweg 4, Delft 2628CD, Netherlands
}%
 
 \author{Rudolf Hanel}
 \email{rudolf.hanel@meduniwien.ac.at}
\affiliation{%
 Institute of the Science of Complex Systems, Center for Medical Data Science, \\ 
 Medical University of Vienna, Spitalgasse 23, 1090, Vienna, Austria
}%
\affiliation{%
 Complexity Science Hub Vienna, \\
 Metternichgasse 8, 1030, Vienna Austria
}%

\date{\today}

\begin{abstract}
    We study self-organization in a minimally nonlinear model of large random ecosystems. Populations evolve over time according to a piecewise linear system of ordinary differential equations subject to a non-negativity constraint resulting in discrete time extinction and revival events. The dynamics are generated by a random elliptic community matrix with tunable correlation strength. We show that, independent of the correlation strength, solutions of the system are confined to subsets of the phase space that can be cast as time-varying Gardner volumes from the theory of learning in neural networks. These volumes decrease with the diversity (i.e. the fraction of extant species) and become exponentially small in the long-time limit. Using standard results from random matrix theory, the changing diversity is then linked to a sequence of contractions and expansions in the spectrum of the community matrix over time, resulting in a sequence of May-type stability problems determining whether the total population evolves toward complete extinction or unbounded growth. 
    In the case of unbounded growth, we show the model allows for a particularly simple nonlinear extension in which the solutions instead evolve towards a new attractor. 
\end{abstract}

\maketitle

\section{Introduction}

Complex systems consisting of many interacting components emerge naturally in the study of ecological communities.
Depending on the context, what these components represent ranges from the populations of animal \cite{hofbauer1998book}, molecular \cite{segel1984book} or microbial \cite{smith1995book} species to groups of individuals in an epidemic \cite{brauer2012book}.
The list of processes governing the interactions between components is enormous, including competition, cooperation, predation, mutualism, and commensalism \cite{harte2011book}.
Dynamical systems used to describe a subset of these processes are often cast either as variants of generalised Lotka-Volterra differential equations \cite{lotka1920,volterra1926} or as (piecewise) linear differential equations.
The latter type arises more specifically in gene-regulatory or biochemical reaction networks, where the components represent molecular or microbial species, such as in the works of Glass-Kauffman \cite{glass1973} and Jain-Krishna \cite{jain1998}; however, (piecewise) linear systems can also  
appear as a linearisation in the neighborhood of an attractor \cite{stokic2008,fedeli2021}.
The central mathematical object in generating the dynamics for both types of system is the community matrix that encodes the interactions between components.
When the ecological community is large and the corresponding community matrix high-dimensional, precise knowledge of the interactions is impossible to obtain.
This has popularized the statistical physics approach to ecology pioneered by Robert May \cite{may1972}, in which the community matrix is assumed to be random. 
This randomness has since been leveraged in countless studies to derive emergent macroscopic properties for the long-time dynamics. For example, phase diagrams \cite{bunin2017}, the typical number of equilibria \cite{ros2023} and the number of surviving species at equilibrium \cite{clenet2023,servan2018}, {as well as abundance distributions for non-equilibrium solutions~\cite{arnoulx2024,mallmin2024}}, have been investigated; for a recent review see also \cite{akjouj2024}.

Independently of the model context, the interacting components often only  have physical significance when their values are nonnegative.
In contrast to Lotka-Volterra systems, linear systems must therefore generally be formulated in a piecewise, discontinuous manner to self-consistently enforce that solutions remain nonnegative at all times \cite{casey2006,jain2001}.
In the random matrix approach, aligning the feasibility constraint with existence criteria for equilibrium solutions is already an interesting problem. 
For Lotka-Volterra systems, classical existence results for globally attracting, non-negative equilibrium solutions require strong dissipativity assumptions on the dynamics \cite{takeuchi1996book}, which translate to strong assumptions on the spectrum of the community matrix~\cite{clenet2022}.
Positive equilibria in particular, where all species coexist, have been shown to require precise fine-tuning of the interactions \cite{grilli2017}.
A mechanism to explain how, nevertheless, large ecological communities stably coexist, is to view them as survivors selected from a much larger species pool, either by a sequence of invasion processes (see \emph{e.g.} \cite{morton1996}) or by initializing the system in a point of coexistence and evaluating the species remaining in the long-time limit \cite{servan2018,bunin2017}.
In the latter case, for Lotka-Volterra systems, the community matrix evaluated at a stable equilibrium is a submatrix of the full set of possible interactions, ``pruned'' by the dynamics.
More specifically, Servan et al. \cite{servan2018} show that the expected number of coexisting species in a globally attracting equilibrium is 1/2 of the total species pool.
In \cite{pettersson2020}, the authors instead consider changes to the community matrix after continuation of the equilibrium in the standard deviation of the interactions. 
A related mechanism of self-organization specific to piecewise linear differential equations was explored in a so-called minimally nonlinear model for chemical reaction networks~\cite{stokic2008}.
In contrast to Lotka-Volterra systems, in this model, species may go extinct in finite time and be revived at a later time through catalytic reactions with the extant species.
The revival mechanism here replaces classical species invasion criteria (see e.g. \cite{akjouj2024}). 
Rather than at an equilibrium and under strong dissipativity assumptions, the submatrix for the remaining active species thus changes over time and can be re-evaluated after each extinction or revival event.
The system was shown to self-organize near an ``inflated edge of chaos'' \cite{hanel2010}, characterized by a broad plateau of vanishing maximal Lyapunov exponent, where always approximately 1/2 of the total species pool remains active.

In this work we revisit the the minimally nonlinear model for random community matrices parametrized by a correlation parameter, $\xi$.
This parameter smoothly interpolates between non-reciprocal ($\xi < 0$) and reciprocal ($\xi > 0$) interactions.
Our focus lies on determining analytically the mechanisms behind the systems ability to self-organize.  
Because the system is piecewise linear, the dynamics can be decomposed into angular and radial variables, measuring the species composition and total population, respectively.
The dynamics of the angular variables determine the sequence of changes made to the community matrix through revival and extinction events, independently of the total population dynamics.
Our main result is that solutions for the angular dynamics are confined to subsets of the phase-space which can be cast as time-dependent ``Gardner volumes'' from the theory of learning in neural networks \cite{gardner1988,amit1989}.
These volumes are shown to decrease with the system diversity (given by the fraction of extant species) and become exponentially small as the diversity approaches a critical threshold.
Numerically we find that solutions evolve to this critical regime in the long-time limit, thereby becoming highly sensitive to perturbations.
While the long-time behavior of solutions depends on the correlation strength, the volumes themselves are independent of this parameter.
When the sequence of extinction and revival events is finite, the community matrix is adapted in such a way that the leading eigenvector becomes both non-negative and strongly localized to a Gardner volume.
This dynamical behavior dominates in the regime of positive correlations.
The radial dynamics are driven by the solution of the angular dynamics and determine whether the system ultimately dies out completely (the origin is a stable equilibrium) or self-sustains (the solution grows without bound).
Using standard results for elliptic random matrices \cite{sommers1988}, we show how the sequence of extinction and revival events causes the spectrum of the community matrix to contract and expand, resulting in a sequence of classical May-like \cite{may1972} stability problems for the origin.
Lastly, the model allows for a particularly simple, fully nonlinear extension which leaves the angular dynamics unchanged. In the extension, a new attractor is created when then origin becomes unstable. Solutions unbounded in the minimally nonlinear model instead converge to this new attractor.

\section{The minimal model} \label{sec:model}

\begin{figure}[t]
 \begin{center}
  \includegraphics[width=1\linewidth]{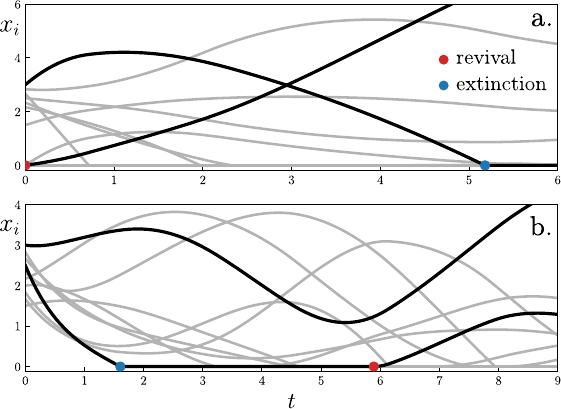}
  \caption{\label{fig:sol_vs_time} Numerical solutions $\bm{x}(t)$ over time for two realizations (a. and b.) of the system~\eqref{eq:MNL} with $d=10$, each with two components $x_1(t),x_2(t)$ highlighted in black. The other components $x_i(t),i>2$ in grey.
  In \textbf{a.} the first component is initially extinct at time $t_0 = 0$ and revived immediately (red dot), while the second, initially extant component, goes extinct at time $t>5$ (blue dot).
  In \textbf{b.} the first initially extant component is driven extinct and subsequently revived at a later time. 
  The second component simply remains active for all time. The index set~\eqref{eq:Ix} changes after each event, as extinct species are removed and revived species added. 
  Numerical integration of~\eqref{eq:MNL} is performed using the explicit Euler method. Extinction events are detected by a positive component becoming non-positive in the next time-step. They are then set to zero.
  Revival events are detected by whether a component at zero becomes positive in the next time-step. 
  }
  \end{center}
 \end{figure}

We consider the dynamics of a large ecosystem comprised of $d$ species, with populations $x_i \ge 0$. The population state is given by points $\bm{x} = (x_1,\dots,x_d)^\top$ in the non-negative cone
\begin{equation}
    C = \{ \bm{x} \in \R^d : x_i \ge 0 \}.
\end{equation}
We define the index set of active (or extant) species in a given population state $\bm{x}$ as
\begin{equation}
\label{eq:Ix}
    I_{\bm{x}} = \{ i : x_i > 0\},
\end{equation}
as well as its complement $I^c_{\bm{x}}$, the index set of inactive (or extinct) species.
The total number of species $d$ in the ecosystem is assumed to be large enough to model their interactions as random.
We construct a network of weighted interactions by drawing a set of $d$ random column vectors
\begin{equation}
\label{eq:kvectors}
    \{\bm{k}_1,\ldots,\bm{k}_d\} \subset \R^d,
\end{equation}
with centered normal $N(0,d^{-1})$ elements $(\bm{k}_i)_j=k_{ij}$ with normalized variance $d^{-1}$.
Together, these define the $d \times d$ community matrix $K=(k_{ij})$.
The positive elements of the community matrix $k_{ij}>0$ represent productive links from species $j$ to species $i$. Negative elements $k_{ij}<0$ instead indicate inhibitory links.
Pairs of entries $(k_{ij},k_{ji})$ for $i<j$ may be correlated with strength $\xi \in [-1,1]$ as 
\begin{equation}
\label{eq:xi}
    \mathbb{E}(k_{ij} k_{ji}) = \frac{\xi}{d},
\end{equation}
where $\mathbb{E}(\cdot)$ denotes expectation with respect to the elements $k_{ij}$.
We will also refer to the correlation strength $\xi$ as the {\em reciprocity}.
It interpolates between symmetric interactions $(\xi=1)$, which are purely reciprocal (competitive, cooperative, etc.) and skew-symmetric interactions ($\xi=-1$), which are purely non-reciprocal (predator-prey, parasitic, consumer-resource, etc.).
The community matrix generates the dynamics of the populations $x_i$ in time according to the following piecewise defined system of ordinary differential equations
\begin{equation}
\label{eq:MNL}
    \frac{d x_i}{dt} = \begin{cases}
        \alpha \bm{k}^\top_i \bm{x} - \beta x_i \quad &\mbox{for } i \in I_{\bm{x}} \\
        \max(0, \alpha \bm{k}^\top_i \bm{x}) \quad &\mbox{for } i \in I^c_{\bm{x}}.
    \end{cases}
\end{equation}
This is the {\em minimally nonlinear model} introduced in \cite{stokic2008}.
It describes the evolution of the populations $x_i$ as a linear dynamical system under the constraints $x_i \ge 0$.
The first of the two cases in equation~\eqref{eq:MNL} describes the dynamics of the extant species $i \in I_{\bm{x}}$.
The second case describes the dynamics of the extinct species, evolving only according to the positive part of the equations for the extant species.
An inactive species is revived when its interactions with the remaining living species are sufficiently cooperative or catalytic, $\bm{k}^\top_i \bm{x} > 0$.
This condition can be viewed as a type of species invasion criterion.
If all components $k_{ij}$ were non-negative, the second case in~\eqref{eq:MNL} would not be necessary to enforce the non-negativity constraint on the components. 
Active species are subject to intrinsic decay or death at a common rate $\beta>0$.
For simplicity, we assume that our ecosystem is closed, in the sense that there is no out- or influx of species by migration or other processes, and that the community matrix is fully connected (cp. \cite{stokic2008}).
Whether or not the ecosystem is capable of self-sustaining or will evolve towards complete extinction is therefore determined by whether or not the dynamics of the total-population can overcome the effect of the depletion terms $-\beta x_i$.
The parameter $\alpha > 0$ scales the standard deviation of the interaction strengths; see also the complexity parameters introduced in \cite{may1972}.

\section{Polar decomposition} \label{sec:decomposition}

\begin{figure}[ht]
 \begin{center}
  \includegraphics[width=.8\linewidth]{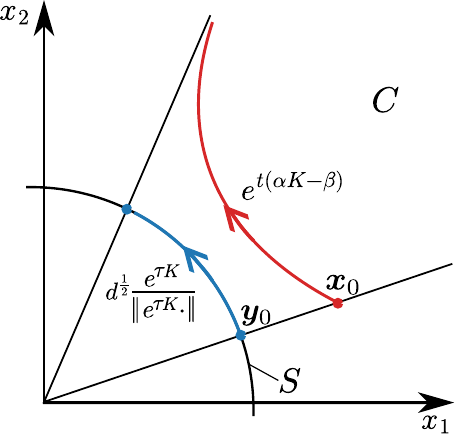}
  \caption{\label{fig:polardecomp} Sketch of the polar decomposition in $\R^2$. The angular variable $\bm{y}$ is the radial projection of $\bm{x}$ through a line through the origin. The projected dynamics depend only on the ordering of the real parts of the eigenvalue. They are independent of the decay rate $\beta$ and the scaling $\alpha$ can be removed by a rescaling of time. The trajectory of the full system \eqref{eq:MNL} is unbounded. The corresponding trajectory for the (compactified) angular dynamics tends to an equilibrium. }
  \end{center}
 \end{figure}

 \begin{figure*}[t]
 \begin{center}
  \includegraphics[width=.83\linewidth]{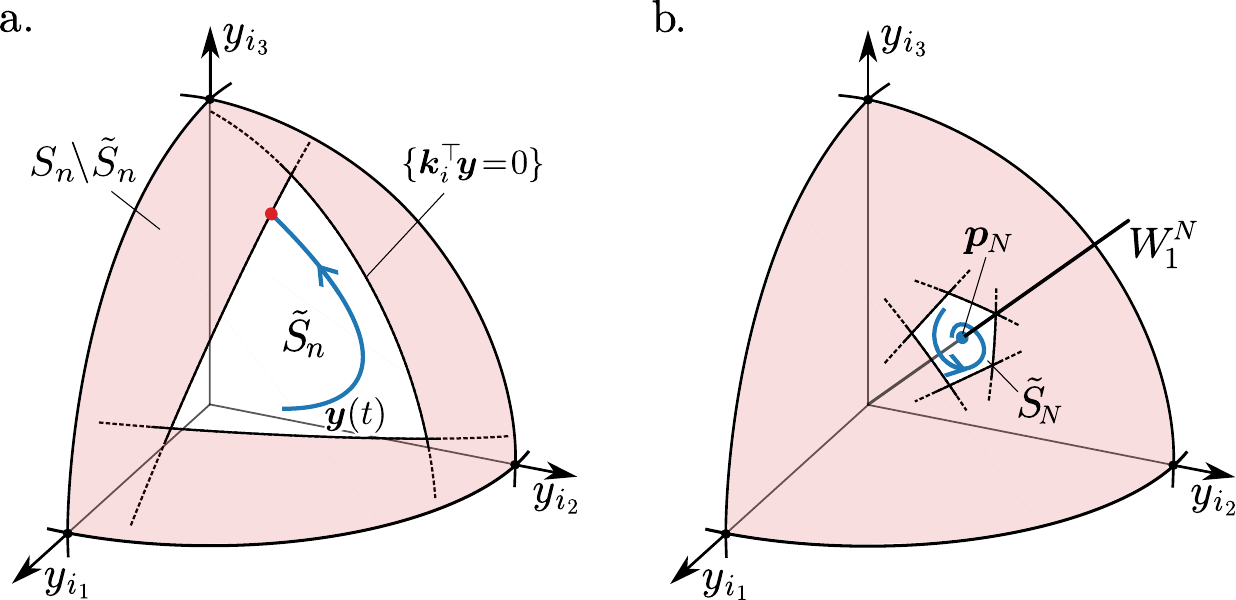}
  \caption{\label{fig:phasespace} Sketch of the decreasing Gardner volumes~\eqref{eq:V} for two values of the diversity $\gamma_n$ and $\gamma_N$ with $\gamma_n > \gamma_N$. \textbf{a.} A solution $\bm{y}(t)$ moves on the set $\tilde{S}_n$ containing the active species $\{i_1,i_2,i_3\} \subset I_n$. The red-dot depicts a revival event where the solution hits any hyperplane $\{ \bm{x} :\bm{k}^\top_i \bm{x} = 0\}$ for $i\in I^c_n$, whose intersections with the sphere are depicted as black lines. \textbf{b.} A solution moves on the set $\tilde{S}_N$ at lower diversity resulting in a smaller Gardner volume. The subset satisfies the feasibility conditions~\eqref{eq:feasibility} and the solution converges to the stable equilibrium $\bm{p}_N$ without further extinction or revival events. }
  \end{center}
 \end{figure*}

A solution $\bm{x}(t)$ of our model~\eqref{eq:MNL} for the initial population state $\bm{x}(0)$ is unique in forward time and absolutely continuous (see Filippov~\cite{filippov1988book}).
Over time, some of the species present in the initial species composition will go extinct or be revived by catalytic interactions with the active species.
As a result, the set \eqref{eq:Ix}, collecting the indices of living species, changes along the solution.
In the following we derive a decomposition of our model~\eqref{eq:MNL}, which allows us to consider $I_{\bm{x}(t)}$ independently from the stability properties of the system.
We define angular and radial components of $\bm{x}$ in the style of Khas'minskii \cite{khasminskii1967}:
\begin{align}
    \bm{y} &=   d^{1/2}\frac{\bm{x}}{\norm{\bm{x}}}, \label{eq:y} \\
   \mbox{and} \quad \rho &= \log \norm{\bm{x}}, \label{eq:rho}
\end{align}
where $\norm{\cdot}$ denotes the 2-norm. 
The variable $\bm{y}$ is the point $\bm{x} \in C\setminus \{0\}$ mapped to the nonnegative section of the sphere of radius $d^{1/2}$
\begin{equation}
\label{eq:S}
    S = \{ \bm{x} \in C : \lVert \bm{x} \rVert^2 = d \},
\end{equation}
by radial projection.
The angular variable $\bm{y}$ is a measure for the species composition.
The radial variable $\rho$ is a measure of the total population. 
Because the population will become exponentially small or large, we consider the logarithm of the norm $\norm{\bm{x}}$.
Differentiation with respect to time of equation~\eqref{eq:y} along the solution $\bm{x}(t)$ yields 
\begin{align}
    \frac{d y_i}{dt} &= d^{1/2} \left( \frac{1}{\norm{\bm{x}}}\frac{d x_i}{dt} - \frac{x_i }{\norm{\bm{x}}^2 } \frac{d }{dt} \norm{\bm{x}} \right)\\
              % &= d^{1/2} \frac{\alpha \bm{k}^\top_i \bm{x} - \beta x_i}{\norm{\bm{x}}} - y_i \frac{ \alpha \sum_j x_j \bm{k}^\top_j \bm{x} -\beta \norm{\bm{x}}^2 }{\norm{x}^2} \\
              &= \alpha ( \bm{k}^\top_i \bm{y} - d^{-1} y_i \sum_j y_j \bm{k}^\top_j \bm{y} ), \quad \mbox{for } \; i \in I_{\bm{x}(t)}. \label{eq:derive_angular}
\end{align}
and
\begin{equation}
    \frac{d y_i}{dt} = \frac{d^{1/2}}{\norm{\bm{x}}} \frac{dx_i}{dt} = \alpha \max(0, \bm{k}^\top_i \bm{y}), \quad \mbox{for } \; i \in I^c_{\bm{x}(t)}.
\end{equation}
These equations are decoupled from the dynamics of the populations $x_i$, up to the criteria that distinguish the dynamics of extinct species from the alive ones in \eqref{eq:MNL} by the index set $I_{\bm{x}(t)}$.
However, we can see from \eqref{eq:y} that $x_i = 0$ is equivalent to $y_i =0$ and that $\bm{k}^\top_i \bm{x} =0$ is equivalent to $\bm{k}^\top_i \bm{y} =0$.
That is, even though the projection~\eqref{eq:y} is not injective, both the extinction and the revival events are mapped one-to-one so that we can replace $I_{\bm{x}}$ with $I_{\bm{y}}$.
Additionally, because the equations are independent of the decay rates $\beta$, we can remove the scaling $\alpha>0$ of the standard deviation by re-parameterizing time as $\tau = \alpha t$ to yield the system of equations
\begin{equation}
\label{eq:angular}
    \frac{d y_i}{d\tau} = \begin{cases}
        \bm{k}^\top_i \bm{y} - L(\bm{y})y_i \quad &\mbox{for } i \in I_{\bm{y}} \\
        \max(0,\bm{k}^\top_i \bm{y}) \quad &\mbox{for } i \in I^c_{\bm{y}},
    \end{cases}
\end{equation}
where we have introduced the scaled quadratic form
\begin{equation}
\label{eq:rayleigh}
    L(\bm{y})= d^{-1} \sum_j y_j \bm{k}^\top_j \bm{y} = d^{-1} \bm{y}^\top K\bm{y}
\end{equation}
Similarly, differentiating equation~\eqref{eq:rho} with respect to time yields a differential equation depending only on the angular variable $\bm{y}$ through the quadratic form
\begin{equation}
\label{eq:radial}
    \frac{d \rho}{d\tau} = L(\bm{y}) - \frac{\beta}{\alpha}.
\end{equation}
In total, we have rewritten the system~\eqref{eq:MNL} on $C\setminus \{0\}$ as the coupled system of equations~\eqref{eq:angular} and \eqref{eq:radial} on the product space $(\bm{y},\rho) \in S \times \R$.
In the new system,  we can first solve the angular dynamics~\eqref{eq:angular} independently, and then use the solution to solve the simple equation for the growth rates $\rho$ on the real line.
In the following, we revert back to using $t = \tau$ as the time variable for the rescaled systems~\eqref{eq:angular} and \eqref{eq:radial}.
The solution $\bm{x}(t)$ of the component dynamics is recovered via the inverse transformation 
\begin{equation}
\label{eq:xinverse}
    \bm{x}(t) = d^{-1/2}\bm{y}(t)\exp(\rho(t)).
\end{equation}
We remark that in
our definition~\eqref{eq:y} of the angular variable, we could have normalized by the $1$-norm and projected to the unit simplex $\{ \lVert \bm{x} \rVert_1^2 = 1\}$ instead of the sphere.
In this case, equation~\eqref{eq:angular} has been studied in \cite{jain2001} as a variant of the continuous time dynamics in the well-known Jain-Krishna model of adaptive networks. 
Our choice of working on the sphere makes the dynamics gradient-like in the special case of symmetric community matrices below.
It also avoids the inconsistency that this version of the Jain-Krishna model does not leave the simplex forward invariant (see \cite[Theorem A.1]{kuehn2019}).
A drawback is that $\rho$ does not directly represent the total population.
The additional scaling by the dimension in the angular variable makes the surface measure of the sphere exponential in $d$, which is convenient for taking large system limits in the following Section.

\section{Angular dynamics: The event sequence}

\subsection{Gardner volumes} \label{sec:gardner}

\begin{figure}[t]
 \begin{center}
  \includegraphics[width=1\linewidth]{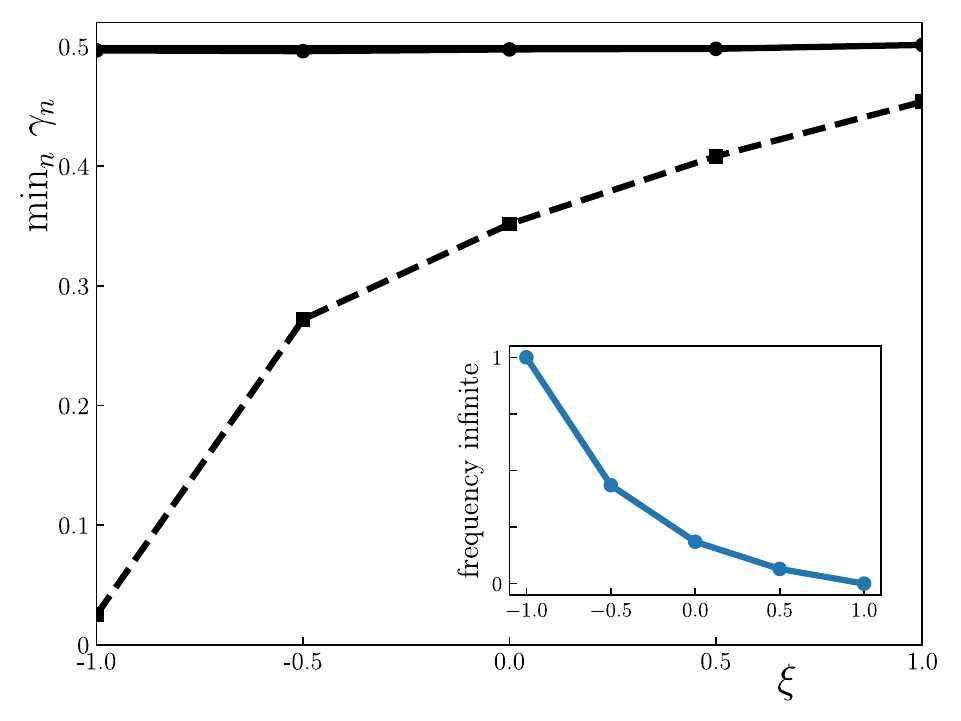}
  \caption{\label{fig:mingamma_vs_xi} Minimal fraction of active species $\gamma_n$ over the course all extinction and revival events for $d = 5000$ averaged over 70 realizations for each value of the correlation strength $\xi = -1,-0.5,0,0.5,1$. The black curve corresponds to the event sequence of the angular dynamics~\eqref{eq:angular}. Solutions which penetrate beyond the critical threshold $\gamma_c = 1/2$ are atypical. The dashed curve corresponds to the model without revival events, see equation~\eqref{eq:MNL_absorbing}. In this case solutions freely move below the critical threshold. Inset shows the empirical probability that an event sequence is infinite for $d=200$. For each value of the correlation strength $\xi$, we average over 200 realizations of the system and count the number of event sequences detected as infinite. For $\xi=+1$, all sequences are finite (see Section~\ref{sec:angular_finite}), and for $\xi=-1$ all sequences are infinite (see Section~\ref{sec:angular_infinite}). An event sequence is detected as infinite, if the time between events $t_{n+1} -t_n$ does not exceed the threshold $500$ and the distance between solution and a leading eigenvector does not drop below $10^{-10}$.
  }
  \end{center}
 \end{figure}

\begin{figure}[ht]
 \begin{center}
  \includegraphics[width=1\linewidth]{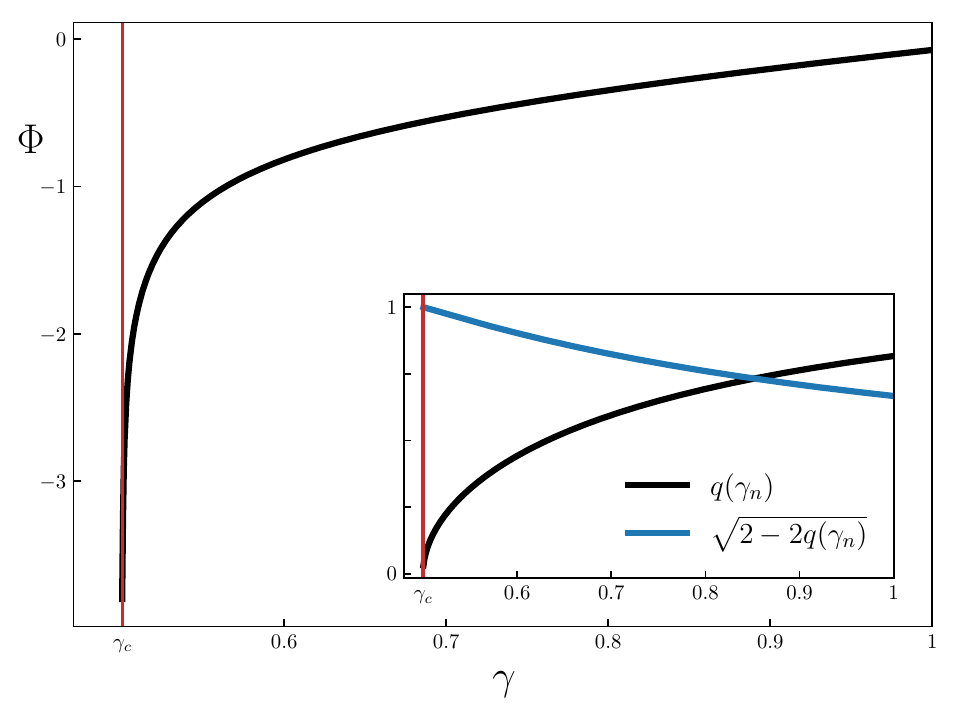}
  \caption{\label{fig:entropy_overlap} The quenched average $\Phi(\gamma_n)$, \eqref{eq:limlogV} determined by numerically solving equation~\eqref{appeq:gardner_entropy}. As the diversity decreases, the fractional volume decreases monotonically. In the limit $\gamma_n \to \gamma_c = 1/2$ of the critical threshold, $\Phi(\gamma_n)$ diverges to $-\infty$. 
  Below the critical threshold, the fractional volume is typically vanishing. Inset shows the typical overlap \eqref{eq:overlap} as function of the fraction of active species, numerically solving equation~\eqref{appeq:qimplicit}. As the number of active species decreases, the overlap monotonically increases towards the limit $q=1$. The corresponding typical distance between solutions \eqref{eq:qdist} plotted in blue. Validity of equation~\eqref{appeq:qimplicit} for the overlap is limit to the regime of large overlaps, but remains in good agreement with also in the case $\gamma_n = 1$ of minimal overlap, see equation~\eqref{appeq:q_noconstraints} in Appendix~\ref{sec:gardner}.}
  \end{center}
 \end{figure}

We consider the independent dynamics of the angular variables $\bm{y}$ given by the system of equations~\eqref{eq:angular}.
Extinction or revival events for a solution $\bm{y}(t)$ occur at discrete times $t_n$.
Ordering this sequence $t_{n}<t_{n+1}$ we can associate with the solution a sequence of active species indices~\eqref{eq:Ix} as
\begin{equation}
\label{eq:event_sequence}
    I_{\bm{y}(t_n)} = I_n, \quad n=0,1,\ldots
\end{equation}
where $t_0 = 0$.
Geometrically, each $I_n$ corresponds to a subset of the phase space $S$ given by
\begin{equation}
\label{eq:Sn}
    S_n = \{ \bm{y} \in S : I_{\bm{y}} = I_n \}.
\end{equation}
When at least one species is extinct, these subsets are the boundary faces of $S$.
We further restrict these subsets to account for revival events as follows:
Using the set of random vectors~\eqref{eq:kvectors} we define the open halfspaces
\begin{equation}
\label{eq:halfspaces}
    H_i = \{ \bm{x} \in \R^d : \bm{k}^\top_i \bm{x} < 0\}, \quad i \in I^c_n.
\end{equation}
According to the model definition~\eqref{eq:angular}, between events $t \in (t_{n},t_{n+1})$ the solution must be contained in the set
\begin{equation}
\label{eq:Stil}
    \tilde{S}_n = S_n \bigcap_{i \in I^c_n} H_i ,
\end{equation}
where the revival condition $\bm{k}^\top_i \bm{y} > 0$ is satisfied for none of the inactive species.
This is the intersection between the boundary face \eqref{eq:Sn} and the convex cone defined by the intersection of all halfspaces \eqref{eq:halfspaces}, see e.g. \cite{cover1967}.
At time $t_{n+1}$ the solution leaves this set as  either
\begin{align}
    y_i(t_{n+1})=0\, &\mbox{ for } i \in I_n \quad \mbox{(extinction)} \label{eq:extinction}\\
    \mbox{or } \; \bm{k}^\top_i \bm{y}(t_{n+1}) = 0\, &\mbox{ for } i \in I^c_n. \quad \mbox{(revival)} \label{eq:revival}
\end{align}
On the other hand, the set of points in $S_n$ inaccessible (or repelling) to the solution is given by the complement $S_n \setminus \tilde{S}_n$,
where the revival condition is satisfied for at least one of the currently inactive species.
Intuitively, when a sufficient number of species is inactive, i.e. the cardinality $|I_n|$ is small, and the vectors $\bm{k}_i$ are not concentrated in some particular direction, then the intersection~\eqref{eq:Stil} is small.   
The domains of the phase space in which we may find a solution for decreasing $|I_n|$ are thus more and more restricted as inequality constraints \eqref{eq:halfspaces} are added; see Figure~\ref{fig:phasespace}.
Taking $\bm{y}$ as a measure of the composition of species in the ecosystem this means that when many species are inactive only a small fraction of compositions remain eligible. 
By leveraging the randomness of the vectors $\bm{k}_i$, this notion can be made precise.
In the following, rather than considering the exact time-ordered sets $I_n$ in \eqref{eq:event_sequence}, we consider their {\em diversity}
\begin{equation}
\label{eq:capacityn}
    \gamma_n = \frac{|I_n|}{d} \in [0,1].
\end{equation}
The fractional volume of $S_n$ that belongs to $\tilde{S}_n$ is given by
\begin{equation}
\label{eq:V}
    V(\gamma_n) = \frac{\text{vol}_{n}(\tilde{S}_n)}{\text{vol}_{n}(S_n)},
\end{equation}
where $\text{vol}_{n}(\cdot)$ denotes the surface measure
of the sphere $\{ \bm{x}' \in \R^{|I_n|}: \norm{\bm{x}'}^2= d \}$.
The fractional volume is a random variable.
Because the inner products~\eqref{eq:halfspaces} reduce to $\bm{k}^\top_i \bm{x} = \sum_{j \in I_n} k_{ij} x_j$ for $i \in I^c_n$, the randomness is determined by the entries of the community matrix for rows corresponding to inactive species and columns corresponding to active species:
\begin{equation}
\label{eq:k_iid}
    k_{ij} \;\mbox{ for } \;(i,j) \in I_n^c \times I_n.
\end{equation} 
Since these pairs of indices are drawn from disjoint sets and correlations~\eqref{eq:xi} arise only for opposing pairs of indices, the components \eqref{eq:k_iid} are independently Gaussian distributed random variables with zero mean for all values $\xi \in [-1,1]$.
Determining $V(\gamma_n)$ can now be cast as a sign-constrained, time-dependent variant of the famous Gardner problem from neural networks and spin-glass physics \cite{gardner1988}.
We are interested in the way in which the fractional volume~\eqref{eq:V}, referred to in the following as the Gardner volume, changes with discrete time events $t_n,n=0,1,\ldots$ and in the critical value $\gamma_c$, such that $V(\gamma_n)$ is typically non-zero for diversities $\gamma_n \ge \gamma_c$, and typically zero for $\gamma_n < \gamma_c$.
By typical we mean here the most-probable value. 
Because the most probable value for the Gardner volume generally does not coincide with the expectation $\mathbb{E}(V(\gamma_n))$, we instead consider the quenched expectation
\begin{align}
\label{eq:limlogV}
    \Phi(\gamma_n) &= \lim_{d \to \infty} d^{-1} \, \mathbb{E}(\log V(\gamma_n))  \\
    &= \lim_{d \to \infty} d^{-1} \, \log V(\gamma_n),
\end{align}
for a fixed value of the diversity $\gamma_n$. 
Here, $\mathbb{E}(\cdot)$ denotes the expectation with respect to the independent entries~\eqref{eq:k_iid}. In \eqref{eq:limlogV} we use that, in the limit $d \to \infty$ the quantity $\Phi(\gamma_n)$ is deterministic, as its variance tends to zero and the most-probable value coincides with the expectation (see e.g. \cite{talagrand1999}).
Adapting a result by Amit, Campbell and Wong \cite{amit1989}, an expression for the above limit can be derived analytically, using the replica-trick
\begin{equation}
    \mathbb{E}(\log V(\gamma_n)) = \lim_{m \to 0^+} m^{-1}  \log \mathbb{E}(V(\gamma_n)^m).
\end{equation}
That is, we first compute the expectation of $m$-copies of the volume for the same realization of the interaction weights \eqref{eq:k_iid} and then take the limits in $m$ and $d$. 
The computation is lengthy, but standard and relegated to Appendix~\ref{app:gardner}.
In the style of the Edwards-Anderson order parameter for spin-glasses \cite{mezard1987book} we can then, for a given $\gamma_n$, determine the fractional volume using the typical overlap 
\begin{equation}
\label{eq:overlap}
    q(\gamma_n) = d^{-1} \mathbb{E}(\bm{y}_1^\top \bm{y}_2) \in [0,1],
\end{equation}
between two points $\bm{y}_{1},\bm{y}_2 \in \tilde{S}_n$ sampled from the remaining phase-space.
In the limit $q(\gamma_n) \to 1^-$ the two points grow closer and the fractional volume tends to zero.
Assuming replica-symmetry, we determine the overlap using the saddle-point approximation~\eqref{appeq:qimplicit}, which can be solved numerically. 
Equivalently, writing the overlap in terms of the Euclidean norm, the typical scaled distance between two points in $\tilde{S}_n$  is
\begin{equation}
\label{eq:qdist}
    d^{-1/2}\norm{\bm{y}_1 - \bm{y}_2} = (2-2q(\gamma_n))^{1/2}.
\end{equation} 
The critical threshold is determined in the limit of large overlaps.
From equation~\eqref{appeq:qimplicit} it is given simply by 
\begin{equation}
\label{eq:capacity_threshold}
    \gamma_c = 1/2,
\end{equation}
i.e. the point $|I_n|=d/2$, where there is an equal number of active and inactive species.
In particular, both the threshold~\eqref{eq:capacity_threshold} and behavior of the volume with $\gamma_n$ are independent of the correlation strength $\xi$ by equation~\eqref{eq:k_iid}.
At the critical threshold $\Phi(\gamma_n)$ diverges to $-\infty$, see Figure~\ref{fig:entropy_overlap}.
We conclude that a solution $\bm{y}(t)$ will not typically spend any open interval of time on subsets $\tilde{S}_n$ for diversities below the critical threshold, $\gamma_n < 1/2$.
This conclusion is supported by numerical results in Figure~\ref{fig:mingamma_vs_xi} showing that the average minimal diversity along a solution does not drop below the threshold for large $d$.
The result also supports the mean-field notion that sequences of subsets selected by the dynamics for a given realization of $K$ and $\bm{y}_0$ do not, on average, deviate significantly in volume from typical Gardner volumes.
Initializing the solution on a subset with cardinality $|I_0|$ below the threshold implies that $\bm{y}_0 \in S_0 \setminus \tilde{S}_0$ and immediately leads to a revival event for a number of inactive species.
Replacing $H_i$ in equation~\eqref{eq:Stil} with the complementary open halfspace $H'_i = \{ \bm{x} \in \R^d : \bm{k}^\top_i \bm{x} > 0\}$ the same argument as above then shows that the number of species revived for such an initial condition is typically less than $d/2$.
The critical threshold~\eqref{eq:capacity_threshold} and the overlap~\eqref{eq:overlap} are central to this paper.
The critical threshold determines the spectral properties of the community matrix $K$ used to decide the stability of the origin with the radial dynamics in Section~\ref{sec:radial}.
The regime of large overlaps is of particular importance when considering the long time dynamics of the angular variables in Sections~\ref{sec:angular_finite} and \ref{sec:angular_infinite}.

\subsection{Finite event sequence} \label{sec:angular_finite}

We show that the long-time behavior for the angular dynamics~\eqref{eq:angular} includes a self-organization process towards community matrices with real leading eigenvalues and non-negative leading eigenvectors.
In particular, the leading eigenvectors are localized to the restricted subsets \eqref{eq:Stil}. 
This self-organization is most prevalent in the case of positive correlation $\xi > 0$.
We begin by considering the solution $\bm{y}(t)$ of the system on the subsets $\tilde{S}_n$ between events.
Here, the system simply behaves as a linear ordinary differential equation projected to the sphere.
Rather than the full community matrix $K$, it suffices to consider the $d \times d$ community matrix $K_n$ simplified to only interactions between species active on $\tilde{S}_n$ given by
\begin{equation}
\label{eq:Kn}
    K_n = \begin{cases}
        (k_{ij}) \quad &\mbox{for } (i,j) \in I_n\times I_n \\
        0 &\mbox{otherwise}.
    \end{cases}
\end{equation}
We denote the set of eigenvalues for the active community matrix $K_n$ by
\begin{equation}
\label{eq:specKn}
    \{ \lambda^n_1,\ldots,\lambda^n_{|I_n|},0,\ldots,0 \},
\end{equation}
where the inactive species combine to a zero-eigenvalue of multiplicity $|I^c_n|$.
We then denote the distinct real parts of the eigenvalues corresponding to the active species by $\sigma^n_i$, ordered as
\begin{equation}
\label{eq:sigma_order}
    \sigma^n_1 > \ldots > \sigma^n_m,
\end{equation}
for $1 \le m \le |I_n|$.
For times $t$ in the interval $(t_n,t_{n+1})$ between two events, the curve $\bm{y}(t)$ is the solution of
\begin{equation}
\label{eq:Oja}
    \frac{d \bm{y}}{dt} = K_n \bm{y} - L(\bm{y}) \bm{y},
\end{equation}
for the initial value $\bm{y}(t_n)$.
It follows that a full solution of \eqref{eq:angular}, starting from any $\bm{y}(t_0) \in S$, is glued together from a sequence of Cauchy problems for the system ~\eqref{eq:Oja} with initial values $\bm{y}(t_n)$ with $n=0,1,\ldots$.
In the following, we will assume that the event sequence is finite.
The solution must then eventually settle in a subset $\tilde{S}_{N}$ without further extinction or revival events for times $t>t_{N}$.
In contrast to a linear differential equation, solutions of~\eqref{eq:Oja} depend only on the ordering of the growth rates~\eqref{eq:sigma_order}, not on their signs.
This can be seen as a consequence of the parameter $\beta$ that determines the sign of the real parts of the eigenvalues, no longer appearing in the factored dynamics \eqref{eq:angular}. 
Therefore, rather than decomposing into center, stable and unstable invariant subspaces, we instead use the ordering~\eqref{eq:sigma_order} to decompose into center, leading and non-leading invariant subspaces (see also \cite[Lemma 4.1.2]{colonius2014book})
\begin{equation}
\label{eq:Rd_decomp}
    \R^d = W^N_0 \oplus W^N_1 \oplus W^N_2.
\end{equation}
Here, $W^N_0$ is the center subspace defined by the zero-eigenvalue of the inactive species and spanned by the set of canonical basis vectors $\{e_i\}$ for ${i \in I^c_{N}}$.
The leading and non-leading subspaces are the direct sums
\begin{equation}
\label{eq:W1W2}
    W^N_1 = \bigoplus_{\real \lambda^{N}_i = \sigma^{N}_1} E^N_i, \quad \mbox{and } \; W^N_2 = \bigoplus_{\real \lambda^{N}_i < \sigma^{N}_1} E^N_i,
\end{equation}
respectively, taken over the real generalized eigenspaces $E^N_i$ of the (complex) eigenvalues $\lambda^{N}_i$.
Each $E^N_i$ is at most two-dimensional because the eigenvalues of the random matrix $K_N$ are almost surely distinct\footnote{up to the zero-eigenvalue corresponding to the inactive species \eqref{eq:specKn}.}.
The leading subspace dominates the long-time behavior of solutions.
To see this, we write the initial condition $\bm{y}(t_{N})$ on $\tilde{S}_{N}$ uniquely as $\bm{y}_N = \bm{w}_1+\bm{w}_2$ for $\bm{w}^N_{1,2} \in W^N_{1,2}$.
From equation~\eqref{eq:y}, for $t>t_{N}$ the solution is given by
\begin{equation}
\label{eq:yt_decomp}
    \bm{y}(t) = d^{1/2} \left( \frac{e^{t K_N} \bm{w}_1}{\norm{e^{t K_N} \bm{y}_N}} + \frac{e^{t K_N} \bm{w}_2}{\norm{e^{t K_N} \bm{y}_N}} \right).
\end{equation}
Initial conditions with $\bm{w}_1 \ne 0$ fill almost the entire set $\tilde{S}_{N}$, the exception being those points contained entirely in the subspace $W^N_2$, where $\text{vol}_N(W^N_2 \cap \tilde{S}_N)=0$.
Asymptotically, $\lVert e^{t K_N} \bm{y}_N \rVert$ must then grow with the exponential leading rate $\sigma^N_1$ and dominate all components evolving in $W^N_2$, with smaller growth rates of at most $\sigma^N_2 < \sigma^N_1$, i.e.
\begin{equation}
\label{eq:convergence_W1}
    \frac{\lVert e^{t K_N} \bm{w}_2 \rVert}{\norm{e^{t K_N} \bm{y}_N}} =O\big(e^{(\sigma^N_2 - \sigma^N_1)t}\big), \quad \mbox{as } t \to \infty.
\end{equation}

It follows that $\bm{y}(t)$ must eventually tend towards the intersection $W^N_1 \cap \cl \tilde{S}_{N}$, where $\cl \cdot$ denotes set closure.
This intersection is non-empty by assumption, as otherwise there would be either an extinction or a revival event for $t>t_{N}$.
We next consider the real subspaces $E^N_i$ determining the leading subspace $W^N_1$ to show that it is one-dimensional.
Supposing that $W^N_1$ contains a number of $E^N_i$ corresponding to conjugate pairs of eigenvalues means $e^{t K_N} \bm{w}_1$ is oscillatory.
The projected solution $d^{1/2} e^{t K_N} \bm{w}_1/\lVert e^{t K_N} \bm{w}_1 \rVert$ on the leading invariant subspace is then either periodic or quasi-periodic, and must leave $\tilde{S}_{N}$ in finite time.
Equations~\eqref{eq:yt_decomp} and~\eqref{eq:convergence_W1} imply that the same eventually holds for the solution $\bm{y}(t)$ which would give another event for $t > t_N$.
Hence $W^N_1$ is the linear subspace spanned by a single real eigenvector $\bm{v}^N_1$ for the eigenvalue $\sigma^N_1$. 
A finite event sequence then implies that the community matrix $K_N$ satisfies the feasibility conditions\footnote{Here we disregard the possibility that the projected eigenvector lies in the boundary $\cl \tilde{S}_{N} \setminus \tilde{S}_N$. In this case the extinction sequence need not be finite (see Section~\ref{sec:angular_infinite}).}.
\begin{equation}
\label{eq:feasibility}
    W^N_1 = \Span \bm{v}^N_1,  \quad \mbox{and} \quad d^{1/2}\frac{\bm{v}^N_1}{\lVert \bm{v}^N_1 \rVert} \in \tilde{S}_{N}. 
\end{equation}
From the previous section, for large $d$ feasibility is further constrained by the condition $\gamma_N \ge 1/2$ on the fraction of active species for large $d$.
For an illustration, see Figure~\ref{fig:phasespace}b.
We denote the projected eigenvector in \eqref{eq:feasibility} by $\bm{p}_N$.
It is a stable equilibrium point for the dynamics
\begin{equation}
\label{eq:convergence}
    \lim_{t \to \infty} \bm{y}(t) =  \bm{p}_N.
\end{equation}

\begin{figure*}[t]
 \begin{center}
  \includegraphics[width=1\linewidth]{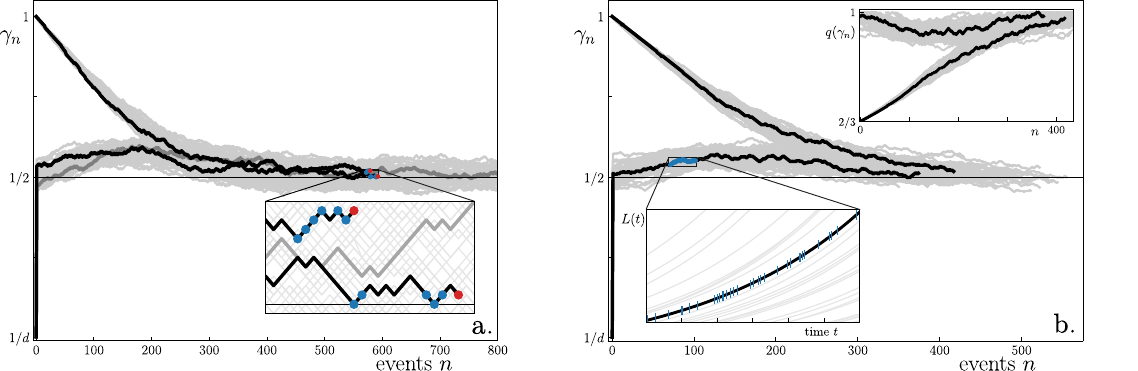}
  \caption{\label{fig:In} Change of the diversity $\gamma_n$ over time (discrete time events $t_n$) for sample solutions for $d=500$ with random initial species compositions $\bm{y}_0$ of fixed cardinality $|I_0|=d$ or $|I_0|=1$. \textbf{a.} The uncorrelated case, $\xi=0$. Event sequences are either finite, when the solution converges to an equilibrium (e.g. the two curves highlighted in black), or infinite (e.g. the curve highlighted in dark grey).
  For $|I_0|=1$ the system immediately undergoes a large revival event towards the critical threshold $\gamma_c = 1/2$.
  Inset shows the final events for the finite sequences. Sets $I_n$ for which $K_n$ satisfies the feasibility conditions~\eqref{eq:feasibility} but undergoes ``accidental'' extinction or revival events marked in blue. The final event beyond which the solution finally converges to an equilibrium marked in red.
  \textbf{b.} The symmetric case, $\xi=1$. All sample solutions converge to an equilibrium with a finite extinction sequence of length $N<600$.
  Lower inset shows the energy $L(t)$ over time along the lower black sample solution across approximately 30 events $n$ marked in blue. While these events include both extinction and revival, the energy remains strictly increasing in time throughout, see~\eqref{eq:gradientlike}. 
  Upper inset shows the change in the overlap $q(\gamma_n) \in [2/3,1]$ over time determined by~\eqref{appeq:qimplicit}.
  The overlap is not defined for values $\gamma_n<1/2$ below the critical diversity threshold.
  }
  \end{center}
 \end{figure*}

More generally, we can associate a stable equilibrium $\bm{p}_N$ to any $\tilde{S}_N$ for which the active community matrix $K_N$ satisfies~\eqref{eq:feasibility} and a corresponding basin of attraction, i.e. the set of all initial points $\bm{y}(t_0) \in S$ convergent to the equilibrium. 
We note that convergence \eqref{eq:convergence} for all initial points mapped to $\tilde{S}_N$ by the dynamics is not guaranteed due to the possibility of ``accidental'' extinction or revival events in the approach to the attractor, see Figure~\ref{fig:In}. 
This may be the case even if the solution arrives on the subset very close to the equilibrium, and when the Euclidean distance to the equilibrium is strictly decreasing in time.
The rate at which non-leading directions decay~\eqref{eq:convergence_W1} is determined by the gap $\sigma^N_1-\sigma^N_2$.
Due to the randomness of the active community matrix, this gap is typically of order $d^{-1/2}$ (see \cite{hoi2015}).
As a consequence, for large $d$, convergence to the equilibrium is increasingly slow.
Lastly, we consider further restriction of the basin of attraction due to the possibility of a basin boundary $W^N_2 \cap \tilde{S}_N \ne \emptyset$, in which case the solution may converge to either $\bm{p}_N$ or to  $-\bm{p}_N$. The latter being unfeasible and leading to a revival or extinction event.
In contrast to the typical overlap \eqref{eq:overlap} between points in $\tilde{S}_N$, the typical overlap between the eigenvectors of the random matrix $K_N$ cannot become arbitrarily large\footnote{This is immediate for normal matrices due to orthogonality, but not when considering eigenvectors of neighboring eigenvalues of non-normal random matrices.} (see \cite{benaych2018} for the case $\xi = 0$). 
Therefore, in the regime of large overlaps $q(\gamma_n)$ we will not find intersections of the eligible phase space $\tilde{S}_N$
with both $W^N_1$ and $W^N_2$.

Numerically, we find equilibria on average only in the regime of large overlaps.
This is in line with results for random Lotka-Volterra systems~\cite{servan2018}, where the expected diversity $\gamma_N$ at an equilibrium is derived to be $\gamma_c$ under diagonal stability assumptions (see also Section~\ref{sec:discussion}).
It is remarkable that by adjusting the active species the system self-organizes into a community matrix $K_N$ with a leading eigenvector not only non-negative, but strongly localized to the subset $\tilde{S}_N$ occupying only a tiny Gardner volume~\eqref{eq:V} of the phase space. 
This also implies that while equilibria may have large basins of attraction, they are sensitive to perturbations.
A solution for the initial value $\bm{y}_0 = \bm{p}_N + \bm{u}_0$ will return directly to the equilibrium for sufficiently small perturbations $\bm{u}_0$ but undergoes revival events when $d^{1/2} \lVert \bm{u}_0 \rVert$ exceeds the typical distance \eqref{eq:qdist}. As $\gamma_n \to \gamma_c$ this distance tends to zero.

\noindent
\\
\textbf{Positive correlation:}
When the solution converges to an equilibrium, the dynamics involve an optimization (selection) process towards the largest available growth rates $\sigma^n_1$.
For an interval $(t_n,t_{n+1})$ of sufficient length, a solution will eventually decrease distance to the leading subspace \eqref{eq:W1W2} for the active species $I_n$.
In the limit of strong positive correlation $\xi \to 1^{-}$ the community matrices $K_n$ are symmetric with real eigenvalues $\sigma^n_i$ and orthogonal eigenvectors $\bm{v}^n_i$.
For a symmetric random matrix we can assume that the leading eigenspace is always one-dimensional and write $W^n_1 = \Span (\bm{v}^n_1)$ and $W^n_2 = \Span(\bm{v}^n_i : i>1)$ for the leading and non-leading subspaces~\eqref{eq:W1W2}.
The optimization process in this case is made explicit by the scaled quadratic form $L(\bm{y})$ defined in equation~\eqref{eq:rayleigh}.
Extreme points of the quadratic form for each $K_n$ are given by the projected eigenvectors $d^{1/2} \bm{v}^n_i/\norm{\bm{v}^n_i}$.
Along a solution $\bm{y}(t)$ for times $(t_n,t_{n+1})$ we have
\begin{align}
    \frac{d}{dt} L(\bm{y}(t)) &= \nabla L(\bm{y})^\top (K_n \bm{y} - L(\bm{y})\bm{y}) \label{eq:dLdt} \\
                &= 2 d^{-1} (\norm{K_n \bm{y}}^2 -  L(\bm{y})^2)
                > 0,
\end{align}
by the Cauchy-Schwarz inequality, assuming that $\bm{y}(t_n)$ is not already contained in $W^n_1$ or $W^n_2$.
Because the quadratic form is also continuous in time across both revival and extinction events we have monotonicity
\begin{equation}
\label{eq:gradientlike}
    L(\bm{y}(t)) < L(\bm{y}(s)), \quad \mbox{ for all } t < s.
\end{equation}
As a consequence, the dynamics in the limit of strong positive correlation are greatly simplified. 
Any initial point $\bm{y}(t_0) \in S$ must converge to an equilibrium point $\bm{p}_N$ on a subset $\tilde{S}_N$ satisfying the feasibility conditions~\eqref{eq:feasibility}.
Generically, extinction sequences are finite\footnote{again disregarding the exceptional case where the equilibrium is contained in the boundary of $\tilde{S}_N$.}.
Monotonicity \eqref{eq:gradientlike} with respect to $L$ does not imply that the event sequence $I_n$ must consist entirely of either revival or extinction events.
It does imply that, if a solution returns to a subset $\tilde{S}_n$ at a later time, it must do so at higher ``energy'', i.e. on the sublevel set 
\begin{equation}
    \{ \bm{y} \in \tilde{S}_n : L(\bm{y}) > L(\bm{y}(t_{n+1})) \},
\end{equation}
For increasing correlation strength $\xi \in [0,1]$, the angular dynamics become more gradientlike \eqref{eq:gradientlike} as opposing pairs of entries $(k_{ij},k_{ji})$ of the community matrix become more symmetric.
We conclude that in the regime of strong positive correlation, equilibrium dynamics dominate and the system consistently self-organizes towards community matrices with non-negative leading eigenvectors strongly localized.
Our numerical results in Figure~\ref{fig:mingamma_vs_xi} indicate that equilibrium dynamics dominate in the entire range $\xi \in [0,1]$.
The ramifications of this on the radial dynamics, determining the overall stability of the system, are explored in Section~\ref{sec:radial}.

\subsection{Infinite event sequence} \label{sec:angular_infinite}

We consider a solution $\bm{y}(t)$ for which the feasibility conditions \eqref{eq:feasibility} are never satisfied for the time-ordered sequence of matrices $K_n$.
This means that the leading subspace $W^n_1$ always contains at least one real eigenspace corresponding to a conjugate pair of eigenvalues.
The argument given in Section~\ref{sec:angular_finite} then shows that the solution leaves each subset $\tilde{S}_n$ in finite time and the event sequence \eqref{eq:event_sequence} is infinite.
The long-time dynamics in this case are richer than in the finite sequence case and we no longer find a self-organizing process restructuring the community matrix in the way of the conditions~\eqref{eq:feasibility}.
In particular, we cannot use the simple asymptotics \eqref{eq:convergence_W1} since the solution need not decrease distance to $W^n_1$ on the interval $(t_n,t_{n+1})$.
Since there are only finitely many possible sets \eqref{eq:Stil}, there must exist an index $N$ such that for $t \ge t_N$ the solution moves between sets 
\begin{equation}
\label{eq:Sinfty}
    \tilde{S}_{N+j}, \quad j=0,\ldots,m-1,    
\end{equation}
each visited infinitely often.
$\bm{y}(t)$ evolves towards a subset of the union of the sets \eqref{eq:Sinfty}. 
The number of events by which the solution may arrive on a given subset scales with the dimension $d$.
We consider the case where the subsets \eqref{eq:Sinfty} form a periodic sequence with minimal period $p \in \mathbb{N}$ such that
\begin{equation}
\label{eq:Sinfty_sequence}
    \tilde{S}_{N+j+p} = \tilde{S}_{N+j}, \quad j =0,\ldots,m-1.
\end{equation}
In the simplest case, it holds that $p=m$ and the solution arrives on a subsets in the sequence always from the same subset by the same event, either extinction~\eqref{eq:extinction} or revival~\eqref{eq:revival}.
Our numerical experiments indicate that traversing a periodic sequence of subsets~\eqref{eq:Sinfty_sequence} results in $\bm{y}(t)$ converging towards a solution $\bm{y}_p(t)$ periodic in time, i.e. $\bm{y}_p(t+T)=\bm{y}_p(t)$ for some minimal period $T>0$.
We find that the diversity $\gamma_{N+j}$ along these stable periodic orbits is near the critical value $1/2$.
As for the equilibria, this implies that the subsets $\tilde{S}_{N+j}$ have small Gardner volumes with the periodic orbit strongly localized and sensitive to perturbations, see equation~\eqref{eq:qdist}.
We note that because solutions of the piecewise defined system are unique only in forward time, a solution periodic in time can be attained in finite time.

\noindent
\\
\textbf{Negative correlation:}
For a random community matrix $K$, we expect the number of eigenvalues on the real axis to decrease with the correlation strength $\xi$ (cp. \cite{byun2023}) in the case of $\xi<0$.
This makes it less likely for the system to attain an active community matrix $K_n$ that satisfies the feasibility conditions~\eqref{eq:feasibility}.
Complementary to the case of strong positive correlation, we can then expect infinite event sequences to dominate in the regime of strong negative correlation. 
In the limit $\xi \to -1^+$, the eigenvalues of any active community matrix are purely imaginary and all event sequences must be infinite.
Numerically we find in this regime a third type of long-time behavior, in which solutions converge to neither an equilbrium nor a period orbit.
Instead they explore large portions of the possible species compositions near the critical diversity $\gamma_c$.
We expect this is the result of $\bm{y}(t)$ being the projection of a quasi periodic solution on each interval $(t_n,t_{n+1})$.
It is interesting to compare at this point the infinite event sequences for strong negative correlations with a variant of the minimal model ~\eqref{eq:MNL}, where the possibility of revival events is removed. 
That is, we consider the angular dynamics
\begin{equation}
\label{eq:MNL_absorbing}
    \frac{d y_i}{dt} = \begin{cases}
        \bm{k}^\top_i \bm{y} - L(\bm{y})y_i \quad &\mbox{for } i \in I_{\bm{y}} \\
        0 \quad &\mbox{for } i \in I^c_{\bm{y}}.
    \end{cases}
\end{equation}
In this system, event sequences consist of only extinction events, are finite, and there is no restriction to the phase space $S_n$ by the intersection of halfspaces~\eqref{eq:Stil}.
The feasibility conditions~\eqref{eq:feasibility} simplify to having a non-negative real leading eigenvector.
The diversity is strictly decreasing in time and the community matrix is pruned as more and more species are driven extinct in order to find such an eigenvector.
As a consequence, when the feasibility conditions are never satisfied, as is the case for $\xi=-1$, the system must collapse entirely.
Numerically we find that the system~\eqref{eq:MNL_absorbing} collapses below the critical threshold~\eqref{eq:capacity_threshold} for all values of $\xi$, see Figure~\ref{fig:mingamma_vs_xi}.
This collapse is difficult to fully resolve numerically for $\xi = -1$, because the oscillations become increasingly slow as the system approaches total extinction.
With revival events, the system has a mechanism for re-diversification when too many species are inactive.
We remark that the event sequence may be infinite even when the feasibility conditions are satisfied at some events $n$ in the sequence due to incidental extinctions described in the previous section. A solution may also converge to an equilibrium in the exceptional case, where this equilibrium is contained in the boundary of a subset.

\begin{figure*}[t]
 \begin{center}
  \includegraphics[width=1\linewidth]{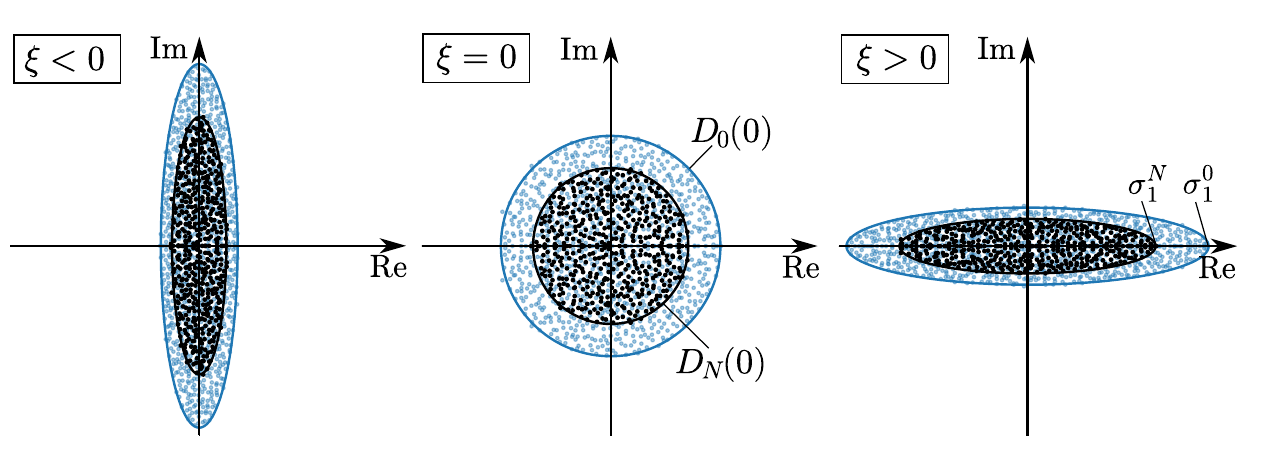}
  \caption{\label{fig:elliptical} Comparison of the spectra of the initial community matrix $K_0 = K$ and the asymptotic community matrix $K_N$ for $d=1000$ for $\xi = -0.65,0,0.65$, each for a single realization of the dynamics.  For the initial community matrix all eigenvalues lie within the blue ellipsoid $D_0$ in the complex plane, up to small errors. After the final extinction event $N$ the spectrum of the active community matrix has contracted to the ellipsoid $D_{N}$ of width of approximately  $2^{-1/2}(1+\xi)$ along the major axis, up to small errors. The contraction of the spectrum in the horizontal direction affecting the leading growth rate $\sigma^N_1$ is largest for strong positive correlation $\xi>0$. Dynamics of the leading growth rates $\sigma^n_1$ depicted in Figure~\ref{fig:fieldofvalues}a.}
  \end{center}
 \end{figure*}

\section{Radial dynamics: The stability problem} \label{sec:radial}

We now consider the dynamics of the growth rates $\rho$ determined by equation~\eqref{eq:radial} as driven by the solution $\bm{y}(t)$ of the angular dynamics.
A solution $\bm{y}(t)$ of \eqref{eq:Oja} defines the simple non-autonomous differential equation on $\R$
\begin{equation}
    \frac{d \rho}{dt} = L(t) - \frac{\beta}{\alpha},
\end{equation}
where $L(t) = L(\bm{y}(t))$.
Integration directly yields the solution
\begin{equation}
    \rho(t) = \rho_0 + \int^{t}_{0} L(s) \,ds - \frac{\beta}{\alpha}t.
\end{equation}
Using the growth rate $\rho$ we determine whether the total norm $\lVert \bm{x} \rVert$ of the system \eqref{eq:MNL} becomes exponentially small (complete extinction) or large (unbounded growth).
While for angular dynamics only the ordering~\eqref{eq:sigma_order} is important, the dynamics of the total norm are decided by whether the growth can overcome the death term $\beta/\alpha$ determined by the intrinsic decay rate $\beta$ and the parameter scaling the standard deviation, $\alpha$.
To consider the stability of the origin for the full system \eqref{eq:MNL}, we define the Lyapunov exponent for the initial value $\bm{x}_0$ as
\begin{equation}
    \lambda(\bm{x}_0) = \limsup_{t \to \infty} t^{-1} \, \rho(t).
\end{equation}
We say that the origin is stable for $\bm{x}_0$, if $\lambda(\bm{x}_0)<0$.
Note that the Lyapunov exponent is independent of the initial total population $\rho_0$.

\subsection{Finite event sequence}

We begin by considering angular dynamics that limit to an equilibrium point $\bm{p}_N \in \tilde{S}_N$ with a finite extinction sequence.
In this case, it suffices to consider only the times $t>t_{N}$ after the final extinction event.
Using that $\bm{p}_N$ is a projected eigenvector for the real leading eigenvalue $\sigma^N_1$ we find
\begin{align}
    \lambda(\bm{x}_0) &= \lim_{t \to \infty} t^{-1} \int^{t}_{t_N} L(s) \, ds - \frac{\beta}{\alpha} \\
                    &  = L(\bm{p}_N) - \frac{\beta}{\alpha}\\
                    & = \sigma^N_1 - \frac{\beta}{\alpha}. \label{eq:lambda_SN}
\end{align}
On the final subset $\tilde{S}_N$ the system is linear and the Lyapunov exponent is simply the real leading eigenvalue of the shifted community matrix $K_N-\beta/\alpha$.
The origin is stable, if $\sigma^N_1 < \beta/\alpha$.
In contrast to a smooth linear system however, the Lyapunov exponent for the minimally nonlinear model is distinctly initial condition dependent. For a given realization of the community matrix, the spectral properties of the final active community matrix $K_N$ depend on which of the (possibly many) feasible equilibria $\bm{p}_N$ the angular dynamics converge to.
In the model context, stability of the origin implies that the system is not able to self-sustain and dies out completely.
We leverage the randomness of the components $k_{ij}$ to compare the spectra of the time-ordered sequence of community matrices $K_n$ for $n=0,1,...,N$.
The stability of the origin, as measured on the traversed subsets $\tilde{S}_n$ by the May-criterion \cite{may1972}
\begin{equation}
\label{eq:sigman-beta}
    \sigma^n_1 < \frac{\beta}{\alpha},
\end{equation}
changes with the discrete time events $t_n$.
The community matrix $K$ defined in Section~\ref{sec:model} is a random matrix with $N(0,d^{-1})$ components with the correlation strength $\xi$ interpolating between the symmetric and skew-symmetric cases.
Its spectral distribution for $\xi \in (-1,1)$ obeys the elliptical law without outliers \cite{nguyen2015}.
By construction, the same holds for non-zero entries of the active community matrix $K_n$ in equation~\eqref{eq:Kn}.
In the limit $d \to \infty$ the distribution of the eigenvalues \eqref{eq:specKn} of the active community matrix, aside from the degenerate zero eigenvalue, is uniform in the subset
\begin{equation}
\label{eq:ellipse}
    D_{n}(\xi)=\Big\{ z \in \mathbb{C} : \frac{\real z^2}{(1+\xi)^2} + \frac{\imaginary z^2}{(1-\xi)^2} \le \gamma_n \Big\},
\end{equation}
of the complex plane \cite{sommers1988}, see Figure~\ref{fig:elliptical}.
The ellipse $D_n(\xi)$ changes over time with the diversity $\gamma_n$.
In particular, the leading growth rate $\sigma^n_1$ closely tracks the semi-axis $(1+\xi) \gamma_n^{1/2}$.
Since the critical threshold $\gamma_c$ \eqref{eq:capacity_threshold} is independent of the correlation strength $\xi$, for large $d$, we get the informal estimate for the leading growth rate
\begin{equation}
\label{eq:stability_sigma1}
   2^{-1/2}(1+\xi) \le \sigma^n_1 \le 1+\xi,
\end{equation}
for all $n=0,1,\ldots,N$ up to small errors.
Here, the upper bound corresponds to $\gamma_n = 1$ where all species are active, and the lower bound corresponds to the critical threshold $\gamma_n = 1/2$ where the maximal number of species are inactive.
For an example, see Figure~\ref{fig:elliptical}.
The Lyapunov exponent is thus determined by the sequence of classical May-like stability problems \eqref{eq:sigman-beta}:
When a sufficient number of species are inactive, the spectrum contracts and the stability of the origin is increased on the interval $(t_n,t_{n+1})$.
Similarly, if a sufficient number of species  are revived, the spectrum expands and the stability of the origin decreases.\footnote{A single extinction (resp. survival) event is certain to contract (resp. expand) the spectrum only for symmetric matrices $\xi = 1$, where Cauchy interlacing inequalities hold.}
For a community matrix $K$ without negative entries (e.g. the adjacency matrix of a graph, \cite{jain1998,horstmeyer2020}), the system~\eqref{eq:MNL} does not undergo extinction events because the non-negative cone $C$ is forward invariant. If we assume that all species are active initially, $K_0 = K$, the origin is unstable for $\beta/\alpha < 1+\xi$ in this case.
By comparison, for our case of centered normal components extinction events can stabilize the origin for the entire range 
\begin{equation}
\label{eq:stability_beta}
    2^{-1/2} (1+\xi) \le \beta/\alpha \le 1+\xi,
\end{equation}
see also Figure~\ref{fig:LE_vs_betac}. 
This stability gap was studied in \cite{stokic2008} as an inflated edge of chaos in terms of the connectance $c$ of a community matrix with random graph structure in the case $\xi=0$. 
This is equivalent to a scaling of the standard deviation $\alpha=c^{1/2}$ in~\eqref{eq:stability_beta}.
\\

\begin{figure}[t]
 \begin{center}
  \includegraphics[width=1\linewidth]{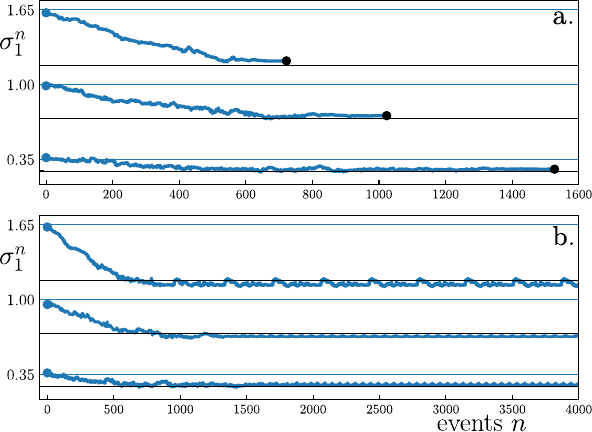}
  \caption{\label{fig:fieldofvalues} Comparison of the leading growth rates in the spectra of the community matrix over time for $K_0 = K$ for $d=1000$ for $\xi = -0.65,0,0.65$, each for a single realization of the dynamics.
  The blue and black horizontal lines correspond to the upper and lower bound on the stability gap~\eqref{eq:stability_sigma1}, respectively.
  \textbf{a.} Finite event sequences with leading growth rate for the final community matrix $K_N$ as the black dot. The full spectrum for $K_0$ and $K_N$ depicted in Figure~\ref{fig:elliptical}. For positive reciprocity $\xi>0$, the initial growth rate is largest but collapses the most under the radial dynamics that determine species composition. For negative reciprocity $\xi<0$ the initial growth rate is smallest but collapses least. 
  \textbf{b.} Infinite event sequences limiting to a periodic sequence of subsets $\tilde{S}_n$ near $\gamma_c$, see equation~\eqref{eq:Sinfty_sequence}, with periodic leading growth rate.  }
  \end{center}
 \end{figure}

In the case of positive correlation $\xi > 0$, the elliptic set is deformed in the horizontal direction.
According to equation~\eqref{eq:stability_beta}, the leading growth rate $\sigma^n_1$ varies more strongly across events for reciprocal interactions, i.e. the stability gap is largest for reciprocal interactions and smallest for non-reciprocal interactions.
Assuming the angular dynamics consistently evolve towards an equilibrium near $\gamma_c$, this does not effect the hierarchy of stability criteria established in \cite{allesina2012}: namely the origin is most likely to be stable for non-reciprocal interactions and least likely to be stable for reciprocal interactions. For an example see Figures~\ref{fig:elliptical} and \ref{fig:fieldofvalues}.

The limit of symmetric matrices $\xi \to 1^-$, yields the interval 
\begin{equation}
    D_{n}(1) = [-2 \gamma_n^{1/2},2\gamma^{1/2}_n ],
\end{equation}
on the real line with spectral distribution obeying the semi-circular law (see e.g. \cite{benaych2011}).

\subsection{Infinite event sequence}

\begin{figure}[t]
 \begin{center}
  \includegraphics[width=1\linewidth]{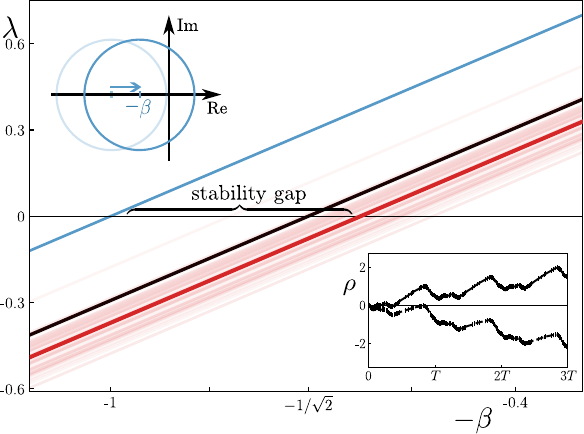}
  \caption{\label{fig:LE_vs_betac} Comparison of the Lyapunov exponent for finite event sequences (black line) and infinite event sequences (red lines) for $\alpha=1$ and $\xi = 0$.
  Each red line corresponds to one realization of the community matrix $K$ and $\bm{y}_0$ for which the event sequence is infinite. For the same realization there are no random fluctuations as $\beta$ is varied because the angular dynamics are independent from it. 
  Decreasing $\beta>0$ corresponds to shifting the spectrum of the community matrix to the right. This is sketched in the top left for $K-\beta$.
  The critical value of the decay rate $\beta$ beyond which the the origin becomes unstable $\lambda(\bm{x}_0)=0$ for finite event sequences is bounded by the value $2^{-1/2}$. The critical value of the decay rate $\beta \approx 0.62 < 2^{-1/2}$ for infinite sequences has been averaged over 60 realizations for $d=100$. Lower inset shows example of a periodic solution $\rho(t)$ for positive (unbounded growth) and negative (complete extinction) Lyapunov exponent.}
  \end{center}
 \end{figure}

We consider angular dynamics with an infinite event sequence for which the solutions limits-to or attains in finite time a periodic solution $\bm{y}_p(t) = \bm{y}_p(t+T)$ with minimal period $T$, contained in the union of subsets~\eqref{eq:Sinfty}. 
The Lyapunov exponent then simplifies to
\begin{align}
    \lambda(\bm{x}_0) &= \frac{1}{T} \int^{T}_{0} L(s) \, ds - \frac{\beta}{\alpha} \\
    & = \frac{1}{T} \sum^{m-1}_{j=0} \int_{t_{N+j}}^{t_{N+j+1}} \bm{y}_p(s)^\top K_{N+j} \bm{y}_p(s) \, ds - \frac{\beta}{\alpha}.
\end{align}
Here, $t_{N+j+1} - t_{N+j}$ is the time spent in the subset $S_{N+j}$ of the phase space in the sequence.
Rather than the May-criterion~\eqref{eq:sigman-beta}, stability of the origin is now determined by the time the solution spends in the set
\begin{equation}
    \{ \bm{y} \in \tilde{S}_{N+j}: \bm{y}^\top K_{N+j} \bm{y} < \beta/\alpha \}.
\end{equation}
on which the total population is strictly decreasing $\frac{d \rho}{dt}<0$. 
In particular, for infinite event sequences the origin may now be stable even when the active community matrix $K_{N+j}$ has positive leading growth rate $\sigma^{N+j}_1-\beta/\alpha > 0$ for all $j = 0,\ldots,m-1$ in the periodic sequence of subsets~\eqref{eq:Sinfty_sequence}.
Numerically we find that, on average, the origin indeed remains stable for infinite sequences for values $\beta/\alpha < 2^{-1/2}(1+\xi)$, see the bold red line in Figure~\ref{fig:LE_vs_betac}.
These results suggest that infinite event sequences have a stronger stabilizing effect on the origin than finite event sequences.

\begin{figure*}[t]
 \begin{center}
  \includegraphics[width=.8\linewidth]{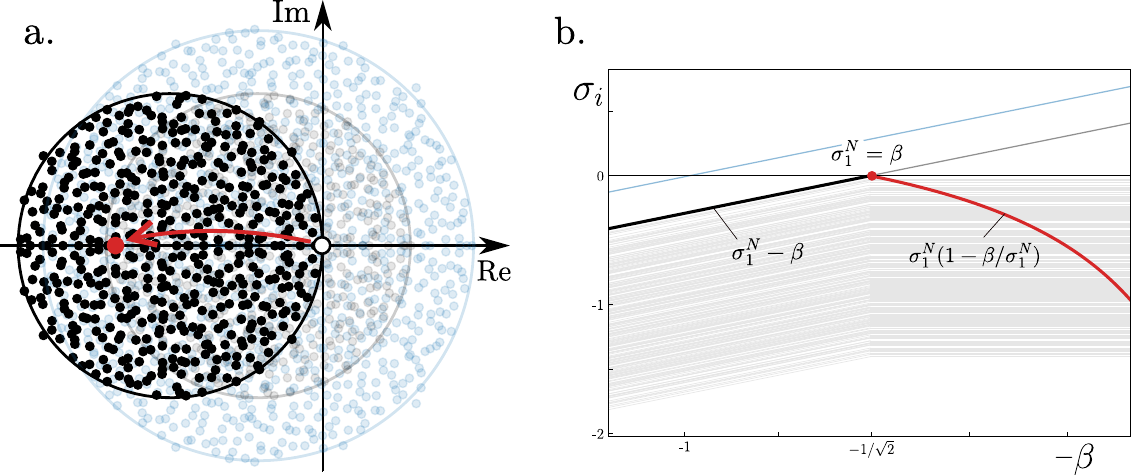}
  \caption{\label{fig:extension} \textbf{a.} Comparison of the spectra of the linearization at the origin at $t=0$ given by $K_0 - \beta$ (in faint blue) and at $t>t_N$ given by $K_N - \beta$ (faint grey), with the linearization at the new equilibrium point~\eqref{eq:spec_xstar} (solid black) for a single realization of $K$. The leading real eigenvalue $\sigma^N_1$ is displaced to $\sigma^N_1(1-\beta/\sigma^N_1)$ inside the bulk spectrum. Parameter values are $d=1000, \alpha = 1$ and $\beta = 0.3$. 
  \textbf{b.} 
   The intrinsic decay rate $\beta > 0$ is varied in the interval $\beta \in [0.3,1.1]$, shifting the spectra. At $\beta=1$ the leading growth rate $\sigma^N_1$ for the full community matrix $K$ crosses through zero. At $\beta=2^{-1/2}$ the origin becomes unstable even for the maximally reduced community matrix $K_N$. Solutions that were unbounded for constant $\alpha$ now converge to the new stable equilibrium $\bm{x}_N$.
  }
  \end{center}
 \end{figure*}

\section{Nonlinear extension} \label{sec:extension}

We briefly consider a very simple nonlinear extension of the model~\eqref{eq:MNL}.
Our goal is to reinterpret the case that the origin is unstable for the radial dynamics discussed in the previous section.
In this case, the solutions $\bm{x}(t)$ of \eqref{eq:MNL} are unbounded.
So far, we have interpreted this case as one in which the interaction structure allows the ecosystem to self-sustain even without an external influx of species.
We now make the dynamics dissipative such that previously unbounded solutions evolve instead towards a new attractor that is different from the origin.
The least complex way to achieve this is to extend the factor $\alpha >0$, scaling the standard deviation, to a smooth function $\alpha(\bm{x})$ on the non-negative cone $C$.
We will consider here the case
\begin{equation}
\label{eq:alphax}
    \alpha(\bm{x}) = 1 - \frac{\norm{\bm{x}}}{A},
\end{equation}
for the carrying-capacity-like number $A>0$.
We view this function $\alpha$ as a type of metabolic scaling. 
It decreases the interaction strengths between species until the intrinsic decay rates $\beta>0$ dominate.
We consider the extended model on the phase space
\begin{equation}
    C' = \{ \bm{x} \in C : \norm{\bm{x}} < A\}.
\end{equation}
The simplicity of the extension lies in the fact that replacing $\alpha \mapsto \alpha(\bm{x})$ in the minimally nonlinear model \eqref{eq:MNL} leaves the angular dynamics unchanged after polar decomposition.
Since the function is radially symmetric, using $\lVert\bm{x}\rVert = \exp(\rho)$ we write $\alpha(\lVert\bm{x}\rVert) = \alpha(\rho)$.
Decoupling the angular dynamics from the radial dynamics now requires a state-dependent re-parametrization of time $\tau_{(\bm{y}_0,\rho_0)} : \R \to \R$ (see e.g. \cite[Chapter 2.2]{katok1995}).
The extinction sequence with and without the minimal extension therefore remains the same.
The radial dynamics in rescaled time are then given by
\begin{equation}
\label{eq:rhodot_extended}
    \frac{d \rho}{dt} = L(\bm{y})- \frac{\beta}{\alpha(\rho)},
\end{equation}
on $\rho \in (-\infty,\log A)$.
For sufficiently large $\rho$, the decay term now dominates and the growth rate is strictly negative.
Solutions $\bm{x}(t)$ for the model~\eqref{eq:MNL} for which the origin was previously unstable, $\lambda(\bm{x}_0)>0$, must now limit to a different attractor in $C'$.
We consider this attractor in the case where $\bm{y}(t)$ limits to an equilibrium point \eqref{eq:convergence} contained in the one-dimensional subspace $W^N_1$.
Restricted to this subspace \eqref{eq:rhodot_extended} has the equilibrium point
\begin{equation}
\label{eq:rhostar}
    \rho_N = \log A \left( 1 - \frac{\beta}{\sigma^N_1} \right).
\end{equation}
In effect, we have replaced the attractor ``at infinity'' for unbounded solutions of the model~\eqref{eq:MNL} with the point $(\bm{p}_N,\rho_N)$ or, using the inverse \eqref{eq:xinverse} the point
\begin{equation}
\label{eq:xN_equilib}
    \bm{x}_N = A \left( 1 - \frac{\beta}{\sigma^N_1} \right) \frac{\bm{v}^N_1}{\norm{\bm{v}^N_1}},
\end{equation}
in the extension.
We see directly that $\rho_N$ is stable for the radial dynamics if the condition
\begin{equation}
\label{eq:stability_xstar}
    \sigma^N_1 > \beta,
\end{equation}
is satisfied, complementary to the stability condition $\sigma^N_1 < \beta/\alpha$ from the previous section.
At the threshold $\sigma^N_1=\beta$, the origin becomes unstable and the new equilibrium \eqref{eq:xN_equilib} is created (a type of transcritical bifurcation of the origin).
Since the radial dynamics are unchanged, from our analysis of the system~\eqref{eq:angular} we know that the point $\bm{x}_N$ is stable with respect to perturbations in the angular directions.
We compare the spectrum of $K_N - \beta$, the linearization at the origin for $t>t_{N}$, with the linearization at the new equilibrium point using the elliptical law \eqref{eq:ellipse}.
The linearization at $\bm{y}_N$ of the dynamics \eqref{eq:Oja} on $\tilde{S}_{N}$ yields the matrix $(K_N - \sigma^N_1)$.
Together with the linearization of \eqref{eq:rhodot_extended}, we find that the spectrum is then given by
\begin{equation}
\label{eq:spec_xstar}
    \{\sigma^N_1(1-\beta/\sigma^N_1), \, \lambda^N_2-\sigma^N_1,\ldots,\lambda^N_{|I_N|}-\sigma^N_1,0,\ldots,0\},
\end{equation}
where the first eigenvalue corresponds to the radial direction, and all other eigenvalues correspond to the angular directions. 
Beyond the transcritical bifurcation, the spectrum of the linearization at the new equilibrium points is thus the spectrum of $K_N$ shifted to lie tangent to the imaginary axis, but with the leading eigenvalue $\sigma^N_1$ shifted into the bulk, see the red arrow in Figure~\ref{fig:extension}.
Because the stability conditions for the origin and \eqref{eq:stability_xstar} are complementary, the stability hierarchy for the new equilibrium is reversed: decreasing the decay term $\beta$ stabilizes $\bm{x}_N$ first for strongly reciprocal and last for strongly non-reciprocal interactions.
The more unstable the origin becomes, the further the eigenvalue $\sigma^N_1(1-\beta/\sigma^N_1)$ for the new equilibrium is pushed into the bulk spectrum, see Figure~\ref{fig:extension}.
From equation~\eqref{eq:rhostar}, the new equilibrium is then located at increasing values of the total population $\rho_N$.
Similarly, in the case where the angular dynamics has an attracting periodic solution $\bm{y}_p(t)$ and the origin is unstable, the extension has a periodic solution $\bm{x}_p(t)$. We expect that, as the intrinsic decay term $\beta$ is decreased, the orbit will be located inside an interval of increasing total population $[\rho_1,\rho_2]$.

\section{Discussion and Outlook} \label{sec:discussion}

It is worth pointing out a few connections between the minimally nonlinear model~\eqref{eq:MNL} and other models in ecology and neural networks theory.
The first is that the angular dynamics \eqref{eq:Oja} can be viewed as a variant of the continuous time dynamics of the Jain-Krishna model for adaptive networks \cite{jain2001}, posed on a sphere rather than the unit simplex (compare Section~\ref{sec:decomposition}). Classically, the Jain-Krishna dynamics assume the community matrix $K$ to be nonnegative \cite{jain1998}, such that the nonnegative cone $C$ is forward invariant and almost all solutions converge to an equilibrium on a subset spanned by the Frobenius-Perron eigenvectors \cite{kuehn2019,horstmeyer2020}. 
For community matrices including also negative entries studied here, the dynamics are richer and limit sets include equilibria and periodic orbits. 
The feasibility conditions~\eqref{eq:feasibility} in this context can be viewed as an extension of the Frobenius-Perron condition.
A distinction from the Jain-Krishna setting is that we also consider the dynamics of the total population. 
Considering only the species composition reduces the dimension of the problem and neatly compactifies the phase-space. However, this approach masks whether a stable attractor for the species composition in fact corresponds to a total population evolving towards complete extinction, see e.g.~\cite{rao2021}. This problem also arises in game-theoretic models, where assuming the number of players remains constant may similarly skew the perspective on sustainability and risks, see e.g. \cite{hauert2006}.
In Section~\ref{sec:extension} we briefly discuss a minimal extension of the model to interpret the case where the population self-sustains and transitions to a new attractor.
Another connection is that, for symmetric community matrices and without sign constraints on the components, equation~\eqref{eq:Oja} describing the angular dynamics between events is known as Oja's learning rule from the theory of neural networks \cite{oja1982}.
Oja's learning rule is also used as a principal component analyzer, since it will converge to the leading eigenvector of the matrix generating the dynamics for almost all initial conditions \cite[Theorem 1]{oja1982}.
In our case, the dynamics instead select for a non-negative leading eigenvector of a submatrix localized to a Gardner volume that depends on the initial condition.
Rather than equation~\eqref{eq:dLdt}, a more elegant way of writing Oja's equation is as a gradient flow on the unit sphere, known as the Rayleigh-quotient gradient flow \cite{mahony2003,helmke2012book}.

The central parameters to our analysis are the reciprocity $\xi$ and the diversity $\gamma_n$.
The reciprocity determines the long-time behavior of solutions favored by the system, while the diversity determines the Gardner volumes for the eligible species compositions.
The volumes are defined by a time-varying number of linear inequality constraints and decrease with  diversity.
Computation of the volumes, as pioneered in \cite{gardner1988}, combines classical results from geometric probability   \cite{wendel1962,cover1965,cover1967,cover1966} with methods from the theory of spin glasses \cite{mezard1987book}.
Linear inequalities appear also in the study of generalised Lotka-Volterra equations and our results here join a growing list of papers similarly combining these fields in ecology; see e.g. \cite{servan2018,grilli2017,cenci2018}. 
We note that the critical threshold $\gamma_c=1/2$ in the minimal model and its independence from the correlation strength is qualitatively similar to results derived in \cite{servan2018}, and that the diversity parameter was previously linked to the stability properties using a similar replica symmetric ansatz in \cite{ros2023}.
We could have derived the critical threshold more directly by adapting a counting theorem due to Cover and Efron \cite[Proposition 1]{cover1966} to our model context.
However, we used the full quenched expectation~\eqref{eq:limlogV} here, because we were interested in explicit changes to the fractional volume and overlap between solutions across events.
By choosing the phase-space for the angular dynamics as a sphere of radius $d^{1/2}$ above, the problem of determining eligible species compositions for a given diversity $\gamma_n$ is transposed into a storage problem in a spherical perceptron (see e.g. \cite{engel2001book}). 
In this context, the random vectors $\bm{k}_i$ defining the halfspaces $H_i$ are a collection of $|I^c_n|$ ``patterns'' to be stored in the neural network using weights drawn from the subset $\tilde{S}_n$.
Rather than $\gamma_n$, the central parameter in the storage problem is called the capacity and given by the ratio of patterns to the dimension $|I_n|$. 
In this analogy, in the minimal model, the number of patterns and thus the difficulty of the storage problem changes over time.
The capacity in this context represents the ratio of extinct to extant species.
In the present work, to make the mapping to the classical Gardner problem as explicit as possible, we consider the minimally nonlinear model with a fully connected community matrix and without external flux or migration terms (cp. \cite{stokic2008}).
With flux terms, the Gardner volumes may become fragmented and non-convex, thereby breaking the assumed replica symmetry, see e.g. \cite{franz2016}.
We believe that, more broadly, the established link to the theory of learning in neural networks may open new interesting avenues of future research on complex ecosystems.

\begin{acknowledgements}
    This research was supported by the European Union’s Horizon 2020 research and innovation programme under the Marie Sklodowska-Curie EvoGamesPlus grant number 955708.
\end{acknowledgements}

\appendix

\begin{widetext}

\section{Gardner volume computation} \label{app:gardner}

We derive expressions for the volume $V(\gamma_n)$ defined in equation~\eqref{eq:V}.
Our arguments follow those in \cite{amit1989}, adapted to the particular setting of our model.
We consider any subset $I_n \subset \{1,\ldots,d\}$ of fixed cardinality $|I_n|=N$ and the diversity
\begin{equation}
    \gamma = \frac{N}{d}.
\end{equation}
The dimension of the problem can be reduced by considering only the subspace of active species.
We redefine the open halfspaces in this setting as 
\begin{equation}
    H_\ell = \{ \bm{x} \in \R^N : \bm{h}_\ell^\top \bm{x} < 0\}
\end{equation}
for the set $\{\bm{h}_1, \ldots, \bm{h}_{d-N}\} \subset \R^N$ of restricted random vectors, where $\bm{h}_\ell=\bm{k}_i\lvert_{I_n}$ denotes the sub-vector with non-zero components indexed by $I_n$.
We then define the measure
\begin{equation}
    \dd \mu(\bm{x}) = \prod_{j=1}^N \frac{\dd x_j}{B} \theta(x_j) \delta(\norm{\bm{x}}^2-d)
\end{equation}
where $\theta(s)$ is the Heaviside function taking the value 1 for positive arguments $s>0$ and zero otherwise.
We include the constant
\begin{equation}
    B = \left(\frac{\pi e}{2 \gamma} \right)^{1/2},
\end{equation}
as a normalization factor to ensure $d^{-1} \log\int \dd \mu(\bm{x}) \to 0$ for $d \to \infty$, keeping fixed the ratio $\gamma$.
That is, $B^N$ is the exponential part of the surface measure of the full positive section of the sphere of radius $d^{1/2}$.
Then we can write the Gardner volume~\eqref{eq:V} as
\begin{equation}
\label{appeq:V}
    V(\gamma) = \int  \dd \mu(\bm{x}) \prod_{\ell=1}^{d-N} \theta(-\bm{h}^\top_\ell \bm{x}).
\end{equation}
The product of the Heaviside-functions yields an indicator function for the intersection of halfspaces $\bigcap^{d-N}_{\ell=1} H_\ell$, while the measure restricts to the non-negative section of the sphere.
To find the average using the Replica trick we first compute the expectation of the product 
\begin{equation}
    \mathbb{E}(V(\gamma)^m) = \int \prod^m_{\varrho = 1} \dd \mu(\bm{x}^\varrho) \; \mathbb{E} \left( \prod^m_{\varrho = 1} \prod_{\ell=1}^{d-N} \theta(-\bm{h}^\top_\ell \bm{x}^\varrho) \right),
\end{equation}
for $m \in \mathbb{N}$. The index $\varrho$ labels the replicas.
We take the expectation with respect to the independent $N(0,d^{-1})$ random variables $h_{\ell j}$ given in equation~\eqref{eq:k_iid}.
We apply the Fourier representation of the Heaviside function
\begin{equation}
    \theta(s) = \int_{[0,\infty)} \dd{\lambda}  \delta( s - \lambda) = \int_{[0,\infty)} \dd{\lambda} \int \frac{\dd{y}}{2 \pi} e^{iy(s-\lambda)}, 
\end{equation}
for each replica $\varrho$. We adopt here the notation that, unless further specified, integral bounds are always over the entire space implied by the integration measure. Then
\begin{equation}
    \mathbb{E} \left( \prod_{\varrho,\ell} \theta(-\bm{h}^\top_\ell \bm{x}^\varrho) \right) = \int_{[0,\infty)^{m(d-N)}} \prod_{\varrho,\ell} \dd{\lambda^\varrho_\ell} \int \prod_{\varrho,\ell} \frac{\dd{y^\varrho_\ell}}{2 \pi}  \mathbb{E} \left( \prod_{\varrho,\ell} e^{i \lambda^\varrho_\ell y^\varrho_\ell + i y^\varrho_\ell \bm{h}^\top_\ell \bm{x}^\varrho  } \right),
\end{equation}
and taking the expectation we find
\begin{align}
    \mathbb{E} \left( \prod_{\varrho,\ell} e^{i \lambda^\varrho_\ell y^\varrho_\ell + i y^\varrho_\ell \bm{h}^\top_\ell \bm{x}^\varrho  } \right) &= \prod_{\ell,j} e^{i \sum_\varrho  \lambda^\varrho_\ell y^\varrho_\ell} \mathbb{E} \left( e^{i h_{\ell j} \sum_\varrho y^\varrho_\ell x^\varrho_j } \right) \\
    &=\prod_{\ell,j} e^{ i \sum_\varrho  \lambda^\varrho_\ell y^\varrho_\ell -\frac{1}{2d} (\sum_\varrho y^\varrho_\ell x^\varrho_j)^2} \\
    & = \prod_\ell e^{ i \sum_\varrho  \lambda^\varrho_\ell y^\varrho_\ell -\frac{1}{2} \sum_\varrho (y^\varrho_\ell)^2 - \sum_{\varrho<\sigma} y^\varrho_\ell y^\sigma_\ell \left( \frac{1}{d} \sum_{j=1}^N x^\varrho_j x^\sigma_j \right) },
\end{align}
where we have used that $\norm{\bm{x}^\varrho}^2 = d$ in the second term of the exponential.
We can now factor the integrals with respect to the $\ell$-index as
\begin{equation}
    \mathbb{E} \left( \prod_{\varrho,\ell} \theta(-\bm{h}^\top_\ell \bm{x}^\varrho) \right) 
    = \left( \int_{[0,\infty)^{m}} \prod_{\varrho} \dd{\lambda^\varrho} \int \prod_{\varrho} \frac{\dd{y^\varrho}}{2 \pi}  e^{ i \sum_\varrho  \lambda^\varrho y^\varrho -\frac{1}{2} \sum_\varrho (y^\varrho)^2 - \sum_{\varrho<\sigma} y^\varrho y^\sigma \left( \frac{1}{d} \sum_{j=1}^N x^\varrho_j x^\sigma_j \right) } \right)^{d-N}.
\end{equation}
Crucially, we then introduce the overlap between replicas as the normalized inner product
\begin{equation}
    q_{\varrho \sigma} =  d^{-1} (\bm{x}^\varrho)^\top \bm{x}^\sigma, 
\end{equation}
taking values between $[-1,1]$ on the sphere of radius $d^{1/2}$ in $\R^N$.
Sampled with respect to the measure $\dd \mu(\bm{x})$ this is the overlap between two points contained in the intersection of the sphere with all halfspaces.
To enforce the variable we insert additional $\delta$-functions as
\begin{equation}
    1 = \int \frac{\dd{q_{\varrho \sigma}} \dd{F_{\varrho \sigma}}}{2 \pi /d} e^{i F_{\varrho \sigma} (d q_{\varrho \sigma} - (\bm{x}^\varrho)^\top \bm{x}^\sigma)}.
\end{equation}
Taking stock, in total we now have
\begin{equation}
    \begin{split}
        \mathbb{E}(V(\gamma)^m) = \int \prod_{\varrho<\sigma} \frac{\dd{q_{\varrho \sigma}} \dd{F_{\varrho \sigma}}}{2 \pi /d} e^{i d \sum_{\varrho < \sigma} q_{\varrho \sigma} F_{\varrho \sigma}} \int \prod_\varrho \dd{\mu(\bm{x}^\varrho)} e^{-i \sum_{\varrho < \sigma} F_{\varrho \sigma} (\bm{x}^\varrho)^\top \bm{x}^\sigma } \\
        \times \left( \int_{[0,\infty)^{m}} \prod_{\varrho} \dd{\lambda^\varrho} \int \prod_{\varrho} \frac{\dd{y^\varrho}}{2 \pi}  e^{ i \sum_\varrho  \lambda^\varrho y^\varrho -\frac{1}{2} \sum_\varrho (y^\varrho)^2 - \sum_{\varrho<\sigma} y^\varrho y^\sigma q_{\varrho \sigma} } \right)^{d-N}
    \end{split}
\end{equation}
In this expression we have coupled the replicas for different indices.
We now want to factor the remaining integrals over the indices $j=1,\ldots,N$, leaving only integrals over the replica indices to then take the limit $m \to 0^+$.
To this end, we must now rewrite our integration measure $\dd{\mu(\bm{x})}$ explicitly in terms of the components $x^\varrho_j$.
Using the integral representation of the $\delta$-function in the integration measure we have
\begin{align}
    \int \prod_\varrho \dd{\mu(\bm{x}^\varrho)} e^{-i \sum_{\varrho < \sigma} F_{\varrho \sigma} (\bm{x}^\varrho)^\top \bm{x}^\sigma } &= \int \prod_{\varrho} \frac{\dd{E_\varrho}}{2\pi} e^{-i d \sum_\varrho E_\varrho} \int_{[0,\infty)^{mN}} \prod_{\varrho,j} \frac{\dd{x^\varrho_j}}{B} e^{i \sum_{\varrho} E_\varrho \norm{\bm{x}^\varrho}^2 - i \sum_{\varrho < \sigma} F_{\varrho \sigma} (\bm{x}^\varrho)^\top \bm{x}^\sigma}, \\
    &= \int \prod_{\varrho} \frac{\dd{E_\varrho}}{2\pi} e^{-i d \sum_\varrho E_\varrho} \prod^N_{j=1} \int_{[0,\infty)^{m}} \prod_{\varrho} \frac{\dd{x^\varrho_j}}{B} e^{i \sum_{\varrho} E_\varrho (x_j^\varrho)^2 - i \sum_{\varrho < \sigma} F_{\varrho \sigma} x^\varrho_j x^\sigma_j} \\
    &=\int \prod_{\varrho} \frac{\dd{E_\varrho}}{2\pi} e^{-i d \sum_\varrho E_\varrho} \left( \int_{[0,\infty)^{m}} \prod_{\varrho} \frac{\dd{x^\varrho}}{B} e^{i \sum_{\varrho} E_\varrho (x^\varrho)^2 - i \sum_{\varrho < \sigma}  F_{\varrho \sigma} x^\varrho x^\sigma} \right)^{N}
\end{align}
where the non-negativity constraint indicator $\prod_j \theta(x_j)$ was used to change the domain of integration to products of $[0,\infty)$.
This leaves
\begin{align}
        \mathbb{E}(V(\gamma)^m) &= 
        \int \prod_{\varrho<\sigma} \frac{\dd{q_{\varrho \sigma}} \dd{F_{\varrho \sigma}}}{2 \pi /d} \int \prod_{\varrho} \frac{\dd{E_\varrho}}{2\pi} \\
        &\qquad \qquad \times e^{d \left( i \sum_{\varrho < \sigma} q_{\varrho \sigma} F_{\varrho \sigma} -i \sum_\varrho E_\varrho \right)} \\
        &\qquad \qquad \times \left( \int_{[0,\infty)^{m}} \prod_{\varrho} \frac{\dd{x^\varrho}}{B} e^{i \sum_{\varrho} E_\varrho (x^\varrho)^2 - i \sum_{\varrho < \sigma}  F_{\varrho \sigma} x^\varrho x^\sigma} \right)^{N} \\
        &\qquad \qquad \times \left( \int_{[0,\infty)^{m}} \prod_{\varrho} \dd{\lambda^\varrho} \int \prod_{\varrho} \frac{\dd{y^\varrho}}{2 \pi}  e^{ i \sum_\varrho  \lambda^\varrho y^\varrho -\frac{1}{2} \sum_\varrho (y^\varrho)^2 - \sum_{\varrho<\sigma} y^\varrho y^\sigma q_{\varrho \sigma} } \right)^{d-N} \\
        &=\int \prod_{\varrho<\sigma} \frac{\dd{q_{\varrho \sigma}} \dd{F_{\varrho \sigma}}}{2 \pi /d} \int \prod_{\varrho} \frac{\dd{E_\varrho}}{2\pi}  \; e^{d \left( G_0(q_{\varrho \sigma},F_{\varrho \sigma},E_\varrho) + \gamma G_1(F_{\varrho \sigma},E_\varrho) + (1-\gamma) G_2(q_{\varrho \sigma}) \right)}. \label{appeq:G0+G1+G2}
\end{align}
In the last line, we have introduced the following functions of the auxilliary parameters $F_{\varrho \sigma},E_\varrho$ and the overlap $q_{\varrho \sigma}$
\begin{align}
    G_0(q_{\varrho \sigma},F_{\varrho \sigma},E_\varrho) &= i \sum_{\varrho < \sigma} q_{\varrho \sigma} F_{\varrho \sigma} -i \sum_\varrho E_\varrho \label{appeq:G0} \\
    G_1(F_{\varrho \sigma},E_\varrho) &= \log  \int_{[0,\infty)^{m}} \prod_{\varrho} \frac{\dd{x^\varrho}}{B} e^{i \sum_{\varrho} E_\varrho (x^\varrho)^2 - i \sum_{\varrho < \sigma}  F_{\varrho \sigma} x^\varrho x^\sigma} \label{appeq:G1}\\
    G_2(q_{\varrho \sigma}) &= \log  \int_{[0,\infty)^{m}} \prod_{\varrho} \dd{\lambda^\varrho} \int \prod_{\varrho} \frac{\dd{y^\varrho}}{2 \pi}  e^{ i \sum_\varrho  \lambda^\varrho y^\varrho -\frac{1}{2} \sum_\varrho (y^\varrho)^2 - \sum_{\varrho<\sigma} y^\varrho y^\sigma q_{\varrho \sigma} }. \label{appeq:G2}
\end{align}
In the limit $d \to \infty$, keeping constant the ratio $\gamma = N/d$, the integral~\eqref{appeq:G0+G1+G2} is dominated by the contribution at the extrema of the functions \eqref{appeq:G0}-\eqref{appeq:G2} with respect to the parameters $(q_{\varrho \sigma},F_{\varrho \sigma},E_\varrho)$ (a saddle-point approximation).
Finding the extrema is greatly simplified by restricting to only those solutions that are replica-symmetric. We make the ansatz:
\begin{equation}
    \begin{cases}
        q_{\varrho \sigma} = q, \quad &\varrho \ne \sigma \\
        F_{\varrho \sigma} = iF, \quad &\varrho \ne \sigma \\
        E_\varrho = iE, \quad &\mbox{for all } \varrho.
    \end{cases}
\end{equation}
Under these assumptions we have
\begin{equation}
    G_0(q,iF,iE)= m \left( \frac{1}{2} q F + E \right) + O(m^2).
\end{equation}
We can also now factor the $\varrho$-integrals for the function~\eqref{appeq:G2} as
\begin{align}
    G_1(iF,iE) &= \log  \int_{[0,\infty)^{m}} \prod_{\varrho} \frac{\dd{x^\varrho}}{B} e^{-\frac{1}{2} (2E+F) \sum_\varrho (x^\varrho)^2 + \frac{1}{2} F (\sum_\varrho x^\varrho)^2} \\
    &= \log \int \frac{\dd{t}}{\sqrt{2\pi}} e^{-\frac{t^2}{2}} \prod_{\varrho=1}^{m} \int_{[0,\infty)} \frac{dx^\varrho}{B} e^{-\frac{1}{2} (2E+F) (x^\varrho)^2 + \sqrt{F} t x^\varrho} \\
    &= \log \int Dt \left( \int_{[0,\infty)} \frac{dx}{B} e^{-\frac{1}{2} (2E+F) x^2 + \sqrt{F} t  x} \right)^m.
\end{align}
Here, we have introduced the standard Gaussian measure $Dt = dt/\sqrt{2 \pi} \, e^{-t^2/2}$ and used the Hubbard-Stratonovich transform to replace the quadratic terms in $x^\varrho$ in the first equation with linear ones.
Completing the square and substitution yields the expression
\begin{align}
    G_1(iF,iE) = -m \log B + \frac{m}{2} \log \frac{2 \pi}{2E+F} + \log \int Dt \, e^{m \frac{F}{2(2E + F} t^2} \left(\int_{[\Omega(F,E)t,\infty)} Dz \right)^m.
\end{align}
Here, we have defined 
\begin{equation}
    \Omega(F,E)= \left(\frac{F}{2E+F}\right)^{1/2}.
\end{equation}
In the limit $m \to 0^+$ this further simplifies at leading order to
\begin{equation}
    G_1(iF,iE) = \frac{m}{2} \left( - 1 + \log 4\gamma  - \log (2E+F) + \frac{F}{2E+F} + 2\int Dt \log H(\Omega(E,F)t) \right) +O(m^2)
\end{equation}
where we have used the constant $B$ to cancel terms corresponding to the surface measure of the sphere. 
We have also introduced the function
\begin{equation}
    H(\omega) = \int_{[\omega,\infty)} Dz.
\end{equation}
The derivation of the function~\eqref{appeq:G2} under replica-symmetry is similar. Because the expression is identical in our setting to the classical result by Gardner~\cite{gardner1988}, see also e.g. \cite{shcherbina2003}, we simply cite it here as 
\begin{equation}
    G_2(q)= m \int D\omega \log H(\Lambda(q)\omega), \quad \Lambda(q) =  \left( \frac{q}{1-q} \right)^{1/2}.
\end{equation}
We would now like to first solve the saddle point equations for $(F,E)$ given by
\begin{align}
    \pdv{F} (G_0(iE,iF,q)+\gamma G_1(iF,iE))&=0 \\
    \pdv{E} (G_0(iE,iF,q)+\gamma G_1(iF,iE))&=0,
\end{align}
as functions of the overlap $q$. 
In contrast to the classical case without the non-negativity constraint, these are not simply algebraic equations due to the remaining integral over the function $H(\Omega t)$ above.
Using $\int Dt \, t \, u(t) = \int Dt\, (d/dt \, u(t))$ partial integration yields the useful relations
\begin{align}
    \pdv{F} \int Dt \log H(\Omega t) &= \frac{-E}{2(2E+F)(E+F)} \int Dt \left[ \frac{\phi(\Omega t)}{H(\Omega t)} \right]^2 \\
    \pdv{E} \int Dt \log H(\Omega t) &= \frac{F}{2(2E+F)(E+F)} \int Dt \left[ \frac{\phi(\Omega t)}{H(\Omega t)} \right]^2,
\end{align}
where $\phi$ denotes a standard Gaussian. 
The integrals can be computed for $\Omega(E,F) \to +\infty$ where the fraction $\phi/H$ has simple asymptotics:
\begin{equation}
\label{appeq:asympt}
    \frac{\phi(\Omega t)}{H(\Omega t)} = \begin{cases}
        O(e^{-(\Omega t)^2/2}), \quad &\mbox{for } t<0 \\
        \Omega t +(\Omega t)^{-1} + O((\Omega t)^{-3})\quad &\mbox{for } t>0
    \end{cases} \;, \quad \mbox{such that} \quad  \int Dt \left[ \frac{\phi(\Omega t)}{H(\Omega t)} \right]^2 = 1+ \frac{\Omega^2}{2} + O(\Omega^{-1}),
\end{equation}
by splitting the integral into positive and negative domains.
Following \cite{amit1989} we expect this limit corresponds to the regime of large overlaps $q$ close to 1. 
Indeed, solving the saddle point equations the leading terms in $1-q$ in the power series are then given by
\begin{align}
    F_0(q) &= \frac{\gamma}{2} (1-q)^{-2} + \frac{3 \gamma}{2} (1-q)^{-1} + O(1), \\
    E_0(q) &= -\frac{\gamma}{4}(1-q)^{-2} + \gamma (1-q)^{-1} + O(1).
\end{align}
Plugging these into equations~\eqref{appeq:G0} and \eqref{appeq:G1} we find 
\begin{equation}
    G_0(q)+\gamma G_1(q)=m \left( \frac{\gamma}{4} (1-q)^{-1} + \frac{3 \gamma}{4} \log (1-q) + O(1) \right),
\end{equation}
where we have again used the asymptotics~\eqref{appeq:asympt} to simplify the integral as $\log H(\Omega t) = \log \phi(\Omega t) - \log \Omega t + O((\Omega t)^{-2})$.
Using this expression we can now write the saddle point approximation for the expectation in the range of large overlaps as
\begin{equation}
\label{appeq:gardner_entropy}
    (d \, m)^{-1} \log\mathbb{E}(V(\gamma)^m) = \min_{q} \left[ (1-\gamma) \int D\omega \log H(\Lambda(q) \omega) + \frac{\gamma}{4} (1-q)^{-1} + \frac{3 \gamma}{4} \log (1-q) \right] + O(d^{-1})
\end{equation}
This expression is similar to the famous Gardner formula \cite{gardner1988}. 
To derive the saddle point equation determining the extremum we again use partial integration to write
\begin{equation}
    \pdv{q} \int D\omega \log H(\Lambda(q)\omega) = -\frac{1}{2}(1-q)^{-1} \int D\omega \left[ \frac{\phi(\Lambda \omega)}{H(\Lambda \omega)} \right]^2,
\end{equation}
such that $q_0$ is the solution of the implicit equation 
\begin{equation}
\label{appeq:qimplicit}
    \gamma -3\gamma (1-q_0) = 2 (1-\gamma)(1-q_0) \int D\omega \left[ \frac{\phi(\Lambda(q_0) \omega)}{H(\Lambda(q_0) \omega)} \right]^2,
\end{equation}
parametrized by $\gamma$.
Note that in equation~\eqref{appeq:gardner_entropy} we have not included any $O(1)$ terms that shift the value of the quenched volume by a constant. These terms do not effect the extremal value of $q$ determined by \eqref{appeq:qimplicit}.
Using the asymptotics~\eqref{appeq:asympt} we take the limit $q \to 1$ in this equation to find the critical threshold given in equation~\eqref{eq:capacity_threshold} of the main text.
The overlap $q$ under replica symmetry is the typical overlap of two points $\bm{x}_1,\bm{x}_2$ sampled uniformly from the Gardner volume~\eqref{appeq:V}
\begin{equation}
\label{appeq:Eoverlap}
    q = \mathbb{E} (d^{-1} \bm{x}^\top_1 \bm{x}_2) = \mathbb{E} \left( \frac{\int \dd \mu(\bm{x}_1) \int \dd \mu(\bm{x}_2) \prod_{\ell=1}^{d-N} \theta(-\bm{h}^\top_\ell \bm{x}_1) \prod_{\ell=1}^{d-N} \theta(-\bm{h}^\top_\ell \bm{x}_2) \; (d^{-1} \bm{x}^\top_1 \bm{x}_2) }{\int \dd \mu(\bm{x}_1) \int \dd \mu(\bm{x}_2) \prod_{\ell=1}^{d-N} \theta(-\bm{h}^\top_\ell \bm{x}_1) \prod_{\ell=1}^{d-N} \theta(-\bm{h}^\top_\ell \bm{x}_2)} \right)
\end{equation}
Computing the typical overlap directly is equivalent to computing the Gardner volume because it appears as the normalization factor (see e.g. \cite[Section 2]{engel2001book}).
The validity of equations~\eqref{appeq:gardner_entropy}-\eqref{appeq:qimplicit} is restricted to a neighborhood of $q=1$ due to our use of the asymptotics \eqref{appeq:asympt}.
Setting $\gamma =1$, i.e. the case without constraints $N=d$, yields $q_0 = 2/3$ in equation~\eqref{appeq:qimplicit}. 
For comparison, we compute in this simple case equation~\eqref{appeq:Eoverlap} directly. Again using a saddle-point approximation yields
\begin{align}
\label{appeq:q_noconstraints}
    q =  \left( \frac{\int \dd \mu(\bm{x}_1) \int \dd \mu(\bm{x}_2)  (d^{-1} \bm{x}^\top_1 \bm{x}_2) }{\int \dd \mu(\bm{x}_1) \int \dd \mu(\bm{x}_2) } \right) = \sum^d_{i=1} d^{-1} \left( \frac{\int \dd \mu(\bm{x}_1) \; x^1_i }{\int \dd \mu(\bm{x}_1)} \right) \left( \frac{\int \dd \mu(\bm{x}_2) \; x^2_i }{\int \dd \mu(\bm{x}_2)} \right) = \frac{2}{\pi} + O(d^{-1}).
\end{align}
Since the value $2/\pi \approx 0.637$ is in good agreement with our result $2/3 \approx0.667$, we assume equation~\eqref{appeq:qimplicit} is a good approximation for the entire range $\gamma \in [0,1]$ throughout the main text.

\end{widetext}

\bibliography{article}

\end{document}